\documentclass[aps, superscriptaddress, secnumarabic, amsmath,amssymb,nofootinbib,nobalancelastpage,tightenlines, a4paper]{revtex4}

\usepackage{eucal} %Makes \mathcal use Euler script instead of the usual Computer Modern calligraphic alphabet; no dependence on amsmath
\usepackage{graphicx}        % standard LaTeX graphics tool
\usepackage{empheq}   %boxes

\usepackage{epstopdf} %essential with TeXShop on Mac

\usepackage{longtable}

\usepackage{esint}  %multiple integral fonts
\usepackage{stmaryrd} %floor, ceiling symbol font
\newcommand{\EParea}{ {\breve{\dot{\sigma}}}}
\newcommand{\halfEParea}{{\breve{\dot{\omega}}}}

\newcommand{\lf}{\llparenthesis}
\newcommand{\rf}{\rrparenthesis}
\newcommand{\lmean}{\lfloor}
\newcommand{\rmean}{\rfloor}

\renewcommand{\vec}[1]{\boldsymbol {#1}}
\newcommand{\tens}[1]{\boldsymbol{\mathsf{#1}}}

\newcommand{\vect}[1]{\vec{#1}}
\newcommand{\F}{\CMcal{F}}
\newcommand{\K}{\mathcal{K}}
\newcommand{\h}{\hbar}
\renewcommand{\path}{C}
\newcommand{\paths}{\mathcal{C}}

\begin{document}

\title{Control Volume Analysis, Entropy Balance and the \\ Entropy Production in Flow Systems}

\author{Robert K. Niven}
\email{r.niven@adfa.edu.au}
\affiliation{School of Engineering and Information Technology, The University of New South Wales at ADFA, Canberra, ACT, 2600, Australia.}
\author{Bernd R. Noack}
\email{Bernd.Noack@univ-poitiers.fr}
\affiliation{Institut PPrime, CNRS -- Universit\'e de Poitiers -- ENSMA, CEAT, 86036 Poitiers Cedex, France.}

\begin{abstract}
This chapter concerns ``control volume analysis'', the standard engineering tool for the analysis of flow systems, and its application to entropy balance calculations. Firstly, the principles of control volume analysis are enunciated and applied to flows of conserved quantities (e.g. mass, momentum, energy) through a control volume, giving integral (Reynolds transport theorem) and differential forms of the conservation equations. Several definitions of steady state are discussed. The concept of ``entropy'' is then established using Jaynes' maximum entropy method, both in general and in equilibrium thermodynamics. The thermodynamic entropy then gives the ``entropy production'' concept. Equations for the entropy production are then derived for simple, integral and infinitesimal flow systems. Some technical aspects are examined, including discrete and continuum representations of volume elements, the effect of radiation, and the analysis of systems subdivided into compartments. A Reynolds decomposition of the entropy production equation then reveals an ``entropy production closure problem'' in fluctuating dissipative systems: even at steady state, the entropy production based on mean flow rates and gradients is not necessarily in balance with the outward entropy fluxes based on mean quantities.  Finally, a direct analysis of an infinitesimal element by Jaynes' maximum entropy method yields a theoretical framework with which to predict the steady state of a flow system. This is cast in terms of a ``minimum flux potential'' principle, which reduces, in different circumstances, to maximum or minimum entropy production (MaxEP or MinEP) principles.  It is hoped that this chapter inspires others to attain a deeper understanding and higher technical rigour in the calculation and extremisation of the entropy production in flow systems of all types.
\\ \\
{\it Reference: Niven, R.K. and Noack, B.R. (2013), Control volume analysis, entropy balance and the entropy production in flow systems, in Dewar R.C., Lineweaver C., Niven R.K., Regenauer-Lieb K., Beyond the Second Law: Entropy Production and Non-Equilibrium Systems, Springer-Verlag, Berlin, Heidelberg, ISBN 978-3-642-40153-4, pp 129-162.}
\end{abstract}

\maketitle

\section{\label{sec:Intro}Introduction}

Over the past half-century, there has been a growing interest in the analysis of non-equilibrium systems -- which by their nature involve flow(s) of one or more quantities -- using variational (extremum) principles based on the rate of thermodynamic entropy production and/or allied concepts. These include the maximum dissipation methods first proposed by Helmholtz \cite{Helmholtz_1868} and Rayleigh \cite{Rayleigh_1913} and their extension to the upper bound theory of turbulent fluid mechanics \cite{Malkus_1956, Busse_1970, Kerswell_2002}; Onsager's ``minimum dissipation'' method \cite{Onsager_1931a, Onsager_1931b}; Prigogine's near-equilibrium minimum entropy production (MinEP) theorem \cite{Prigogine_1967, Kondepudi_P_1998}; the far-from-equilibrium maximum entropy production (MaxEP) principle advocated by Paltridge \cite{Paltridge_1975, Paltridge_1978}, Ziegler \cite{Ziegler_1977} and others \cite{Ozawa_etal_2003, Kleidon_L_book_2005, Martyushev_S_2006, Bruers_2007c}, the main focus of this book; a MinEP {framework for} engineering design advocated particularly by Bejan \cite{Bejan_1996}; a MinEP limit on transitions between equilibria \cite{Salamon_A_G_B_1980, Salamon_B_1983, Nulton_etal_1985} %Andresen_G_1994, Crooks_2007} 
or steady states \cite{Niven_Andresen_2009} respectively in thermodynamic or flow systems; and various minimum and maximum power methods applied to electrical circuits \cite{Kondepudi_P_1998, Jeans_1966, Landauer_1975, Jaynes_1980, Zupanovic_etal_2004, 
Christen_2006}
%Botric_etal_2005, %, Bruers_etal_2007a} 
and pipe flow networks \cite{Thomas_1942, PaulusJr_G_2004, Martyushev_2007, Niven_2010_JNET}. A broader category of variational technique consists of the maximum relative entropy (MaxEnt) method of Jaynes \cite{Jaynes_1957, Jaynes_1963, Jaynes_2003, Tribus_1961a, Tribus_1961b}, which has seen myriad applications in many fields \cite{Kapur_K_1992} and has been used in efforts to explain the above MaxEP / MinEP principles \cite{Dewar_2003, Dewar_2005, Niven_2009_PRE, Niven_2010_PTB, Niven_2012_AIP}. Such a zoo of different variational principles provides considerable scope for confusion, especially given their competing claims and partisanship. The entropy concept itself -- and in consequence the thermodynamic entropy production -- also provides a fertile ground for misunderstanding, which never ceases to yield unexpected traps for beginners and (even) well-established researchers. 

In engineering, the method of {\it control volume analysis} is generally regarded as the most important tool for the analysis of flow systems, underpinning virtually all vehicular, fluid transport, energy generation, manufacturing, civil infrastructure and environmental control systems, and whose basic principles apply to all flows \cite{Prager_1961, Aris_1962, Tai_1992, Street_etal_1996, Spurk_1997, Munson_etal_2010}.  Recently, the authors have been surprised by the lack of appreciation of the control volume method throughout the sciences, even in those disciplines which -- one would think -- might gain the most from their use.  For example, both an ``ecosystem'' and a ``soil'' are control volumes, which experience various material and energy flows (inputs and outputs) through their boundaries, and which undergo various internal processes. Their mathematical modelling therefore requires careful control volume analysis.  Indeed, although not commonly calculated by engineers, the concept of entropy production itself arises from a control volume analysis of a dissipative system, and can be fruitfully examined from this perspective.

The aim of this chapter is to clarify the basis of the entropy production concept of non-equilibrium thermodynamics -- and in consequence its extremisation -- using the principles of control volume analysis. In \S\ref{sec:CV}, the control volume method and its main results are presented, and applied to flows of various quantities, for both integral and differential forms.  Several definitions of steady state are then discussed. In \S\ref{sec:genericH}, we examine the (generic) entropy concept (here labelled $\mathfrak{H}$), which in turn reduces, by a Jaynes' MaxEnt analysis of an equilibrium system, to the thermodynamic entropy $S$ (\S\ref{sec:S}). Control volume analysis of the latter (\S\ref{sec:EB}) enables rigorous definitions of the total thermodynamic entropy production $\dot{\sigma}$ and its local form $\hat{\dot{\sigma}}$. Several special features of the entropy balance are examined, including discrete and continuum representations, radiative effects, compartmentalisation and the definition of steady state.  In \S\ref{sec:closure},  a Reynolds decomposition is used to reveal an ``entropy production closure problem'', manifested as a discrepancy between the overall mean and mean-of-products components. Finally, in \S\ref{sec:FlowSys} we analyse an infinitesimal control volume by Jaynes' MaxEnt method to directly predict the steady state. This yields a theoretical framework which reduces to (secondary) MaxEP or MinEP principles in different circumstances.  The main motivation for  this chapter is to inspire others to attain a deeper understanding and higher technical rigour in the calculation and extremisation of the entropy production in flow systems of all types. 

\section{\label{sec:CV}Justification and Principles of Control Volume Analysis}

\textbf{Two Descriptions:} Historically, two approaches have been developed for the analysis of flow systems \cite{Prager_1961, Aris_1962, Tai_1992, Street_etal_1996, Spurk_1997, Munson_etal_2010}:
\begin{enumerate}
\item{The {\it Lagrangian description}, which follows the behaviour of individual particles (either molecules or infinitesimal fluid elements) as they move, and so examines individual trajectories within the flow; and} 
%in physics, usually molecules; in fluid mechanics, usually infinitesimal fluid elements
\item{The {\it Eulerian description}, which examines particular points or regions in space through which the flow passes, and so considers the flow field. }
\end{enumerate}
The Lagrangian approach has attained a high prominence in physics, giving rise to the field of classical mechanics (e.g.\ equations of motion, action integrals, principle of least action, Hamiltonian function, Liouville's theorem) and the concept of position-momentum phase space \cite{Lanczos_1970}. It also provided the basis of 19th century statistical physics, including Maxwell's velocity distribution, Boltzmann's H-theorem and their successors (including modern lattice-Boltzmann methods) \cite{Harris_1971}, and of 20th century stochastic analyses, such as Markov processes and the Fokker-Planck and Master equations \cite{Risken_1984}. 
%Lagrangian methods have also been applied almost exclusively to certain non-fluid flows, such as electromagnetic radiation \cite{**}.  
For all this prominence, however, Lagrangian methods impose considerable computational difficulties and are not widely used in engineering practice, except in specific cases where their use becomes essential (e.g. early re-entry of spacecraft through rarefied gases).  Instead, the vast bulk of engineering fluid flow, heat and mass transfer calculations are conducted using the Eulerian description, necessitating a control volume analysis.  

%The classical analysis of flow systems by the phase space concept also exhibits another difficulty: the non-conservation of phase space in a dissipative system, which voids Liouville's theorem \cite{**}. In other words, flow systems cannot be simply characterised by the positions $\vec{x}$ and momenta $\vec{p}$ of their constituent particles, but require additional parameters associated with correlations between these parameters and the degradation of motion into heat.  The application of Lagrangian methods to dissipative systems thus requires a high degree of care, and may become computationally intractable unless some other principle can be invoked.

\textbf{Control Volume Analysis:} We now introduce the engineering concept of a {\it control volume} (CV), a geometric region through which one or more fluid(s) can flow, surrounded by a well-defined boundary or {\it control surface} (CS). The control volume is assumed to be embedded within a surrounding {\it environment} (or ``rest of the universe'') which maintains the flow(s).  We also require the concept of a {\it fluid volume} (FV) (in some references a {\it material volume} \cite{Prager_1961} or {\it system} \cite{Street_etal_1996, Munson_etal_2010}), an identifiable body of fluid particles (or differential ``fluid elements'') which moves with time, bounded by its {\it fluid surface} (FS). We therefore analyse the motion of a fluid volume through a control volume. 

\begin{figure}[t]
\begin{center}
\setlength{\unitlength}{0.6pt}
  \begin{picture}(500,175)
   \put(0,0){(a)}
   \put(0,10){\includegraphics[width=55mm]{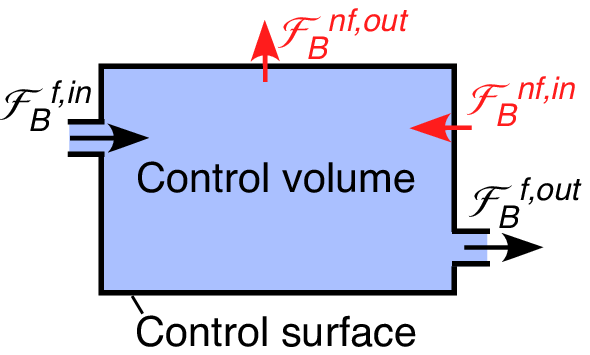}}
   \put(300,0){(b)}
   \put(300,10){\includegraphics[width=45mm]{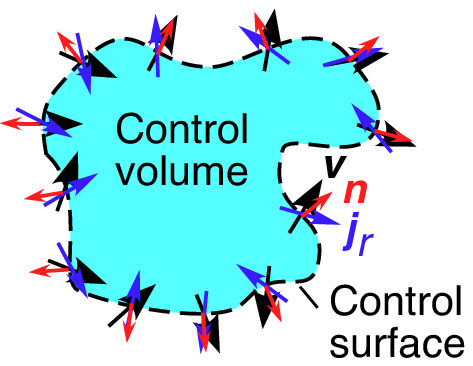}}
  \end{picture}
\end{center}
\caption{Example control volumes for the analysis of (a) simple (flow rate) and (b) integral (vector flux) flow systems, showing representative fluid and non-fluid flow parameters.}
\label{fig:CV}
\end{figure}

Consider the simple fixed, non-deforming control volume shown in Figure \ref{fig:CV}(a), which experiences a discrete set of time-varying flow rate(s) across its control surface, and may also undergo various time-varying rate processes within its volume. We also consider the fluid volume coincident with the control volume at time $t$, which migrates downstream to a different position at time $t + dt$. For each conserved quantity $B$ (e.g. mass, energy, momentum), the rates of change of $B$ within the fluid and control volumes are connected by the conservation equation \cite{Prager_1961, Aris_1962, Tai_1992, Street_etal_1996, Spurk_1997, Munson_etal_2010}:
\begin{equation}
\boxed{
\frac{DB_{FV(t)}}{Dt} = \frac{\partial B_{CV}}{\partial t} + \F_{B,f}^{out} - \F_{B,f}^{in} 
}
\label{eq:Bbalance} 
\end{equation}
where $DB_{FV(t)}/Dt$ is the {\it substantial}, {\it material} or {\it total derivative} of $B$, denoting its rate of change in motion with the fluid; ${\partial B_{CV}}/{\partial t}$ is the rate of change of $B$ within the control volume\footnote{Strictly, for a fixed and non-deforming control volume, this should be written ${dB_{CV}}/{dt}$. The partial derivative is adopted to avoid confusion with some authors' use of ${dB_{CV}}/{dt}$ to denote the substantial derivative, and for consistency with broader applications to moving control volumes.}; and $\F_{B,f}^{out}$ and $\F_{B,f}^{in}$ are respectively the outward and inward flow rates of $B$ due to fluid flow through the control surface\footnote{In engineering, it is standard practice to designate flow rates by an overdot, here $\dot{B}$. In deference to the different meaning of the overdot in physics, to signify a rate of production within a system, $\F_{B}$ is used herein for a bulk flow rate of $B$.}. In \eqref{eq:Bbalance}, the flow rates only refer to fluid-borne flows; all other flows of $B$ are accounted within the substantial derivative ${DB_{FV(t)}}/{Dt}$.  Note the ``out -- in'' form of \eqref{eq:Bbalance}: in many texts it is written in the opposite sense (often in different notation):
\begin{equation}
\frac{\partial B_{CV}}{\partial t} = \frac{DB_{FV(t)}}{Dt} + \F_{B,f}^{in} - \F_{B,f}^{out} 
\label{eq:Bbalance2} 
\end{equation}
but the meaning is identical.  If we understand the processes by which $B$ changes within its fluid volume (both internal and external), their rate of change can be equated to ${DB_{FV(t)}}/{Dt}$, yielding an overall balance equation for $B$.

Now consider the more complicated geometry of Figure \ref{fig:CV}(b), in which the flow of $B$ is represented by its time-varying fluid-borne flux $\rho b \vec{v}$ (measured in SI units of $[B]$ m$^{-2}$ s$^{-1}$) through the control surface, where $\rho(\vec{x},t)$ is the fluid density, $b(\vec{x},t)$ is the  specific (per unit fluid mass) density of $B$ and $\vec{v}(\vec{x},t)$ is the local (mass-average) velocity, in which $\vec{x}$ denotes position and $t$ time. The $B$ balance equation becomes:
\begin{equation}
\frac{DB_{FV(t)}}{Dt} =\frac{\partial B_{CV}}{\partial t}  + \oiint\limits_{CS} \rho b \vec{v} \cdot \vec{n} dA 
\label{eq:Re_transp_flux} 
\end{equation}
where $\vec{n}(\vec{x} \in \text{CS})$ is the unit normal to the control surface (positive outwards), $A$ is the surface area and $\oiint\nolimits_{CS}$ denotes integration around the control surface.  Expressing $B_{CV}=\iiint\nolimits_{CV} \rho b dV$, where $V$ is the volume, \eqref{eq:Re_transp_flux} reduces to \cite{Prager_1961, Aris_1962, Tai_1992, Street_etal_1996, Spurk_1997,  Munson_etal_2010}:
\begin{equation}
\boxed{
\frac{DB_{FV(t)}}{Dt} 
%=\frac{\partial B_{CV}}{\partial t}  + \oiint\limits_{CS} \rho b \vec{v} \cdot \vec{n} dA 
= \frac{\partial }{\partial t} \iiint\limits_{CV} \rho b dV  + \oiint\limits_{CS} \rho b \vec{v} \cdot \vec{n} dA
}
\label{eq:Re_transp} 
\end{equation}
Eq.\ \eqref{eq:Re_transp} is known as Reynolds' transport theorem. 

Since the control volume used here is stationary and non-deforming, the partial derivative in \eqref{eq:Re_transp} can be brought inside the integral.  Furthermore, from Gauss' divergence theorem, $\oiint\nolimits_{CS} \rho b \vec{v} \cdot \vec{n} dA = \iiint\nolimits_{CV} \nabla \cdot (\rho b \vec{v}) dV$, %where $\nabla \cdot$ is the divergence, 
so \eqref{eq:Re_transp} can be written:
\begin{equation}
\frac{DB_{FV(t)}}{Dt} = \iiint\limits_{CV} \biggl[  \frac{\partial }{\partial t} \rho b  +   \nabla \cdot (\rho b \vec{v}) \biggr] dV
\label{eq:Re_transp_div} 
\end{equation}
Also, by integration over mass elements $dm = \rho dV$ of the fluid mass $M$ \cite{Spurk_1997}:
\begin{equation}
%\begin{split}
\frac{DB_{FV(t)}}{Dt}
=\frac{D}{Dt} \iiint\limits_{FV(t)} \rho b dV 
=\frac{D}{Dt} \int\limits_{M} b dm
= \int\limits_{M} \frac{Db}{Dt} dm
= \iiint\limits_{FV(t)} \rho \frac{Db}{Dt}  dV 
%\end{split}
\label{eq:Re_transp_mass}  %mass integral
\end{equation}
using the local substantial derivative $Db/Dt = \partial b/ \partial t + \vec{v} \cdot \nabla b$. 

%\textcolor{blue}{**Need to make true for any CV; otherwise extra f'l term
%\\
%- Panton "Incompr Flows" - 3 types of volumes
%}

Eqs.\ \eqref{eq:Re_transp_div}-\eqref{eq:Re_transp_mass} are valid for fluid and control volumes of any size, including infinitesimal volumes $dV$.  It is therefore permissible to equate their integrands, assuming coincident fluid and control volumes in the infinitesimal limit, to give a differential conservation equation for each element $dV$ in the fluid \cite{Spurk_1997, Durst_2008}: 
%any of \cite{Prager_1961, Aris_1962, Tai_1992, Street_etal_1996, Munson_etal_2010, **} ?? - don't think so
%any of Kuiken_1994, Demirel_2002 ?
%\footnote{From \eqref{eq:Re_transp_mass}, the left term of \eqref{eq:cons_diff} given by \cite{Steeter_etal_1998}, p196, is in general incorrect.}: 
\begin{equation}
\boxed{
%\frac{D}{Dt}(\rho b)_{FV(t)} =  
\rho \frac{Db}{Dt}
%= \rho \frac{\partial b}{\partial t} + \rho \vect{v} \cdot \nabla b  
=  \frac{\partial }{\partial t} \rho b  +   \nabla \cdot (\rho b \vec{v})
%=  \frac{D}{D t} (\rho b)  +   \rho b \nabla \cdot \vect{v}
}
\label{eq:cons_diff} 
\end{equation}  
The left-hand term can be further equated to the sum of rates of change of $\rho b$ in the infinitesimal fluid volume, due to internal and external processes, giving a local balance equation for $B$.  As with all local formulations, \eqref{eq:cons_diff} employs the {\it continuum assumption}, in which the system is assumed much larger than the molecular scale, so that its behaviour can be considered continuous even in the infinitesimal limit \cite{Prager_1961}.  Eqs.\ \eqref{eq:Re_transp} and \eqref{eq:cons_diff} represent two long-standing traditions of fluid mechanics, integral (global) and local conservation laws, for the analysis of flow systems. 

The particular forms of \eqref{eq:cons_diff} for seven physical quantities are listed in Table \ref{tab:cons}.  Here ``$\cdot$'' is the vector scalar product, ``$:$'' is the tensor scalar product, $^{\top}$ is a vector or tensor transpose, %$\vec{x}$ the position vector;   
 $[\vec{\delta}, \vec{\epsilon}]$ are the Kronecker delta and third-order permutation tensors; %$N_c$ 
$[\rho_c, n_c, M_c, \vec{j}_c, \hat{\dot{\xi}}_c]$ are respectively the %moles, 
mass density, molar density (molality), molar mass, molar flux and molar rate of production of species $c$; $[P, \vec{\tau}, \psi]$ are the pressure, stress tensor (positive for compression) and mass-weighted potential; 
$[\vec{g}_c, \psi_c]$ are the specific body force and potential on species $c$; and %$E, E_M$
$[e, e_M, u, \vec{j}_Q]$ are the %total energy, kinetic + potential energy, 
specific total energy, specific kinetic + potential energy, specific internal energy and heat flux. All fluxes $j_Q$ and $j_c$ are measured relative to the local mass-average fluid velocity $\vec{v}$. 
%; and $[z, \rho_z, \vec{j}_z, \hat{\dot{\xi}}_z]$ are the charge, charge density, charge flux and rate of production of charge. 
The listed equations are valid for compressible flow under fairly broad assumptions, assuming conservative body forces $\vec{g}_c = - \nabla \psi_c$ on each species $c$. Other formulations can be derived for different circumstances \cite{deGroot_M_1984, Kuiken_1994, Demirel_2002}.

\begin{table}[t]
%%\arrayrulecolor{black}
\caption{Seven differential balance equations \eqref{eq:cons_diff} for compressible flow (adapted after \cite{Durst_2008, deGroot_M_1984, Bird_etal_2006}).}
\label{tab:cons}       % Give a unique label
%
% Follow this input for your own table layout
%
\begin{tabular}{p{3.0cm}p{2.2cm}p{6.5cm}}
\hline\noalign{\smallskip}
Property $B$ &$b$ &Balance equation (differential form)\\
\hline\noalign{\smallskip}
%%\arrayrulecolor{white}
Fluid mass  &1	
&  
$0 
= \frac{\partial }{\partial t} \rho  +   \nabla \cdot (\rho \vec{v}) $ 
\\
%\hline\noalign{\smallskip}
%{Species mass}  &$m_c$ &
%$\frac{m_c}{m} = \frac{\rho_c}{\rho}$ 
%&%$\frac{D}{Dt} \rho_c =  
%$\rho \frac{D }{Dt} \bigl( \frac{\rho_c}{\rho} \bigr)
%= \frac{\partial }{\partial t} \rho_c  +   \nabla \cdot (\rho_c \vec{v}) 
%= - \nabla \cdot \vec{j}_c + \hat{\dot{\xi}}_c$
%\\
\hline\noalign{\smallskip}
Species moles  &
$\frac{N_c}{m} = n_c$ 
&  
$\rho \frac{D n_c}{Dt} 
= \frac{\partial }{\partial t} \rho n_c  +   \nabla \cdot (\rho n_c \vec{v}) 
= - \nabla \cdot \vec{j}_c + \hat{\dot{\xi}}_c$
\\
\hline\noalign{\smallskip}
Linear momentum   &$\vec{v}$
&  
$\rho \frac{D \vec{v}}{Dt} 
= \frac{\partial }{\partial t} (\rho \vec{v})  +   \nabla \cdot (\rho \vec{v} \vec{v}^\top)
= -\nabla P - \nabla \cdot \vec{\tau} + \sum\nolimits_c \rho_c \vec{g}_c$
\\
\hline\noalign{\smallskip}
Angular  &$\vec{x} \times \vec{v}$
& 
$\rho \frac{D }{Dt} (\vec{x} \times \vec{v})
=  \frac{\partial }{\partial t} \rho (\vec{x} \times \vec{v})  +   \nabla \cdot \rho \vec{v} (\vec{x} \times \vec{v})$ 
\\
momentum &&\hspace{3pt} $=-\nabla \cdot ( \vec{x} \times P \vec{\delta})^\top - \nabla \cdot (\vec{x} \times \vec{\tau})^\top + \sum\nolimits_c (\vec{x} \times \rho_c \vec{g}_c) - \vec{\epsilon} : \vec{\tau} $
\\
\hline\noalign{\smallskip}
Total energy  &$e=e_M + u$
& 
$\rho \frac{De}{Dt} =  \frac{\partial }{\partial t} (\rho e)  +   \nabla \cdot (\rho e \vec{v})$ \\
&&\hspace{3pt} $= - \nabla \cdot \vec{j}_Q - \nabla \cdot (P \vec{v}) - \nabla \cdot (\vec{\tau} \cdot \vec{v}) - \sum\nolimits_c M_c \nabla \cdot (\psi_c \vec{j}_c)$
\\
\hline\noalign{\smallskip}
Kinetic + potential  &$e_M$
&  
$\rho \frac{D e_M}{Dt} 
= \frac{\partial }{\partial t} (\rho e_M)  +   \nabla \cdot (\rho e_M \vec{v})$ 
\\
energy &$ = \frac{1}{2} |\vec{v}|^2 + \psi$&\hspace{3pt} 
$= - \vec{v} \cdot \nabla P  -  \vec{v} \cdot (\nabla \cdot \vec{\tau})  - \sum\nolimits_c M_c \psi_c \nabla \cdot \vec{j}_c$
\\
\hline\noalign{\smallskip}
Internal energy  &$u$
& 
$\rho \frac{Du}{Dt} =  \frac{\partial }{\partial t} (\rho u)  +   \nabla \cdot (\rho u \vec{v})
$ \\&&\hspace{3pt} $
= - \nabla \cdot \vec{j}_Q - P \nabla \cdot  \vec{v} - \vec{\tau} : \nabla \vec{v}  - \sum\nolimits_c M_c \vec{j}_c \cdot \nabla \psi_c$
\\
%\hline\noalign{\smallskip}
%{Charge} &$z$ &$\frac{z}{m} = \frac{\rho_z}{\rho}$
%&  
%$\rho \frac{D}{Dt} \bigl( \frac{\rho_z}{\rho} \bigr)
%= \frac{\partial }{\partial t} \rho_z  +   \nabla \cdot (\rho_z \vec{v}) = - \nabla \cdot \vec{j}_z +  \hat{\dot{\xi}}_z$
%\\
%%\arrayrulecolor{black}
\noalign{\smallskip}\hline\noalign{\smallskip}
\multicolumn{3}{l}{
Assumptions and relations: }
%NOTE: Kuiken 1994 gives quite different eqns for lin + ang momentum, on species basis !!
\\
\multicolumn{3}{l}{
%KEEP** (a) $\sum\nolimits_c n_c M_c =1$,  $\sum\nolimits_c n_c M_c \vec{v}_c=\vec{v}$, $\sum\nolimits_c \vec{j}_c M_c = 0$ and $\vec{j}_c = \rho n_c (\vec{v}_c - \vec{v})$, with $n_c M_c$ = mass fraction of $c$.
(i) $\rho_c = \rho n_c M_c$, $\sum\nolimits_c \rho_c \vec{v}_c = \rho \vec{v}$, $\sum\nolimits_c n_c M_c =1$,  $\sum\nolimits_c n_c M_c \vec{v}_c=\vec{v}$, $\sum\nolimits_c \vec{j}_c M_c = 0$ and $\vec{j}_c = \rho n_c (\vec{v}_c - \vec{v})$.
} \\
\multicolumn{3}{l}{
(ii) $\hat{\dot{\xi}}_c = \sum\nolimits_d \chi_{cd} \hat{\dot{\xi}}_d$ and $\sum\nolimits_c \chi_{cd} = 0$. 
%} \\
%\multicolumn{4}{l}{
\hspace{35pt}
(iii) $\vec{g}_c = - \nabla \psi_c$, $\rho \psi = \sum\nolimits_c \rho_c \psi_c$ and $\sum\nolimits_c \psi_c \chi_{cd}=0$.
} \\
\noalign{\smallskip}\hline\noalign{\smallskip}
\end{tabular}
%$^a$ Table foot note (with superscript)
\end{table}

\textbf{Steady State:} %
We now define ${\partial B_{CV}}/{\partial t}=0$ as the {\it stationary} or {\it steady state} of a control volume. From \eqref{eq:Re_transp}-\eqref{eq:Re_transp_div}:
\begin{align}
\frac{\partial B_{CV}}{\partial t}=0  
\hspace{15pt} &\Rightarrow \hspace{15pt} 
\frac{DB_{FV(t)}}{Dt} \biggr|_{st} 
= \oiint\limits_{CS} \rho b \vec{v} \cdot \vec{n} \, dA
=\iiint\limits_{CV} \nabla \cdot (\rho b \vec{v}) \; dV
\label{eq:Re_transp_stst} 
\end{align}
where $st$ denotes steady state.  We see that at steady state, the internal change of quantity $B$ within the fluid volume is exactly balanced by its flux out of the control surface, and hence its integrated divergence. Similarly, using \eqref{eq:cons_diff} and the definition of divergence \cite{Tai_1992, Kreyszig_1993}, we can define the steady state for an infinitesimal element:
\begin{align}
\frac{\partial }{\partial t} \rho b = 0 
\hspace{15pt} &\Rightarrow \hspace{15pt} 
%\frac{D(\rho b)_{FV(t)}}{Dt} \biggr|_{st} 
\rho \frac{Db}{Dt} \biggr|_{st}
= \lim\limits_{CV \to 0} \frac{ \oiint\limits_{CS} \rho b \vec{v} \cdot \vec{n} dA}{\iiint\limits_{CV} dV}   %DEF OF DIVERGENCE
= \nabla \cdot (\rho b \vec{v})
\label{eq:cons_diff_stst} 
\end{align}
Since both $\vec{v}(\vec{x},t)$ and $B_{FV(t)}$ (or $b(\vec{x},t)$) are time-dependent, a steady state can involve time-varying fluxes, provided these are exactly balanced by time-varying internal changes.  In practice, however, any variability in the fluxes and/or rates will render \eqref{eq:Re_transp_stst}-\eqref{eq:cons_diff_stst} almost impossible to achieve (we could call them a {\it strict steady state}). It is therefore common in fluid mechanics (but not stated explicitly) to consider the {\it mean steady state} ${\partial \langle B \rangle_{CV}}/{\partial t}=0$, where $\langle B \rangle$ denotes some mean (stationary first central moment) of $B$, referred to as a Reynolds average \cite{Spurk_1997, Durst_2008, Schlichting_2001, White_2006}. Usually, $\langle B \rangle$ is equated with the time mean $\overline{B} =\lim\nolimits_{T \to \infty} T^{-1} \int\nolimits_{0}^T B dt$. In some situations, the ensemble mean $\widetilde{B} =\lim\nolimits_{K \to \infty} K^{-1} \sum\nolimits_{k=1}^K B^{(k)}$ is used, where $B^{(k)}$ is the $k$th realisation of $B$ \cite{Spurk_1997}. For the latter, it is {usual practice} to invoke the {\it ergodic hypothesis}, in which the ensemble mean is assumed equivalent to the time mean; this assumption is correct only for certain types of flows.  From \eqref{eq:Re_transp} and \eqref{eq:cons_diff}:

%\textcolor{blue}{**disting between stat stationary and non-stat flows
%\\
%Monin \& Yaglom}

\begin{empheq}[box=\fbox]{align}
%\boxed{
\frac{\partial \langle B \rangle_{CV}}{\partial t}=0  
\hspace{10pt} &\Rightarrow \hspace{10pt} 
\frac{D\langle B \rangle_{FV(t)}}{Dt}  
= \oiint\limits_{CS} \langle \rho b  \vec{v} \rangle \cdot \vec{n} dA
=\iiint\limits_{CV} \nabla \cdot \langle \rho b \vec{v} \rangle dV
\label{eq:Re_transp_stst2} 
\\
\frac{\partial }{\partial t} \langle \rho b \rangle = 0 
\hspace{10pt} &\Rightarrow \hspace{10pt} 
%\frac{D \langle \rho b \rangle_{FV(t)} }{Dt}  \biggr|_{\langle st \rangle} 
 \biggl \langle \rho \frac{Db}{Dt} \biggr \rangle 
%= \lim\limits_{CV \to 0} \frac{ \oiint\limits_{CS} \langle \rho b \vec{v} \rangle \cdot \vec{n} dA}{\iiint\limits_{CV} dV}   %DEF OF DIVERGENCE - IN MEAN
=     \nabla \cdot \langle \rho b \vec{v} \rangle
%}
\label{eq:cons_diff_stst2}  
\end{empheq}
These give much more useful definitions than \eqref{eq:Re_transp_stst}-\eqref{eq:cons_diff_stst}\footnote{In consequence, the mean steady state need not be steady! Indeed the Fluctuation Theorem provides a strong argument that, far from equilibrium, it cannot be steady \cite{Niven_2010_PTB}.}. 
Importantly, since $\langle B \rangle_{CV} =\iiint\nolimits_{CV} \langle \rho b \rangle dV$ for a stationary control volume, the global and local mean steady states \eqref{eq:Re_transp_stst2}-\eqref{eq:cons_diff_stst2} are equivalent, provided both are measured over long time periods.  In contrast, the global and local strict steady states \eqref{eq:Re_transp_stst}-\eqref{eq:cons_diff_stst} are not equivalent, except for time-invariant fluxes and internal processes at both global and infinitesimal scales. 

%A third approach to the steady state, which differs from the mean, is given with reference to the entropy production in \cite{**}. 

Throughout this chapter, the term {\it equilibrium} is used exclusively in its thermodynamic sense, to indicate the stationary state of a thermodynamic system, while {\it steady state} (usually qualified) refers to the stationary state of a control volume.  

\textbf{Further Remarks:} %
Control volume analysis thus provides a rigorous framework for the analysis of flow systems, but like all mathematical methods, it holds some traps for beginners.  Firstly, it is {\it essential} that the control volume and its control surface be clearly defined. This almost always requires a schematic diagram. Different control volumes represent different systems (with different steady states) and in general will yield different results. Where is the control surface? Which flows actually pass through the boundary and so must be included? Which flows are internal and so can be neglected? This study also considers only stationary control volumes. A moving and/or deforming control volume may be advantageous in some circumstances, but requires additional care \cite{Prager_1961, Aris_1962, Tai_1992, Munson_etal_2010}. 
Finally, if a control volume is compartmentalised into sub-volumes, each of which is analysed by balance equations \eqref{eq:Bbalance} or \eqref{eq:Re_transp}, the geometry of each compartment must be clearly defined, so that all flows can be identified and attributed to the correct compartments and external or internal boundaries. 

\section{\label{sec:Entropy}Concept of Entropy}
\subsection{\label{sec:genericH}Generic (Information) Entropy}

We now turn to the entropy concept, which causes many difficulties but in actual fact is very simple. While many justifications are available, arguably the most profound is the combinatorial basis expounded by Boltzmann and Planck \cite{Boltzmann_1877, Planck_1901}, in which we seek the {\it most probable state} of a probabilistic system.  The system is typically represented by an allocation scheme in which $N$ entities (balls) are distributed amongst $I$ categories (boxes), forming individual {\it microstates} or {\it configurations} of the system.  These are then grouped into observable {\it macrostates} or {\it realizations} of the system, specified by the number of balls $n_i$ in each $i$th box. For distinguishable balls and boxes, the probability of a specified realization is given by the multinomial distribution:
\begin{equation}
\mathbb{P} = Prob({n_1,...,n_I}|N,{q_1,...,q_I}) = N! \prod\limits_{i=1}^I \frac{q_i^{n_i}}{n_i !}
\label{eq:multinomial}
\end{equation}
where $q_i$ is the prior or source probability of a ball in the $i$th box or, in other words, its assigned probability before observation.
Seeking the maximum of $\mathbb{P}$, we recognise (as did Boltzmann \cite{Boltzmann_1877}) that it is easier to maximise %$\ln \mathbb{P}$:
%\begin{align}
$\ln \mathbb{P} 
= \ln N! + \sum\nolimits_{i=1}^I ( n_i \ln q_i  - \ln n_i!)$. 
%\label{eq:multinomial2}
%\end{align}
Introducing the Stirling approximation $\ln N! \approx N \ln N - N$ in the asymptotic limit $N \to \infty$ (or alternatively the Sanov \cite{Sanov_1957} theorem), with some rearrangement we obtain:
\begin{equation}
\boxed{
%\begin{split}
\mathfrak{H} 
= \lim\limits_{N \to \infty} \; \frac{1}{N} \ln \mathbb{P} 
%=   - \sum\limits_{i=1}^I  \frac{n_i}{N} \ln \frac{n_i/N}{q_i}   
%\\
=  - \sum\limits_{i=1}^I p_i \ln \frac{p_i}{q_i}   
%\end{split}
}
\label{eq:KL}
\end{equation}
where we take $p_i = \lim\nolimits_{N \to \infty} \, n_i/N$ as the actual (observed or {\it a posteriori}) probability of a ball in the $i$th box. The function $\mathfrak{H}$ is referred to as the {\it relative entropy} or (negative) Kullback-Leibler function \cite{Kullback_L_1951}.  For equal priors $q_i = I^{-1}$, this simplifies to the {\it Shannon entropy} \cite{Shannon_1948}:
\begin{equation}
%\begin{split}
\lim\limits_{N \to \infty} \; \frac{1}{N}\ln \mathbb{P}_{\text{equal } q_i} 
\cong 
\mathfrak{H}_{Sh} 
= - \sum\limits_{i=1}^I p_i \ln {p_i} 
%\end{split}
\label{eq:Shannon}
\end{equation}
modulo a constant.  Provided the system is indeed multinomial \eqref{eq:multinomial}, maximising the relative entropy \eqref{eq:KL} (or Shannon entropy \eqref{eq:Shannon} for equal $q_i$), subject to any constraints, gives the most asymptotically probable realization of the system.  

Adopting this probabilistic (or combinatorial) basis of entropy, we see that Jaynes' MaxEnt method \cite{Jaynes_1957, Jaynes_1963, Jaynes_2003} can be applied to {\it any} probabilistic system, not just in thermodynamics. For maximisation, it is necessary to incorporate the normalisation constraint and (usually) $R$ moment constraints, respectively:
\begin{gather}
\sum\limits_{i = 1}^I {p_i }= 1, 
\hspace{10pt} \text{and} \hspace{13pt}
\sum\limits_{i = 1}^I {p_i f_{ri} }= \langle {f_r } \rangle, \quad r = 1,...,R, 
\label{eq:Cr}
\end{gather}
where $f_{ri}$ is the $i$th value of property $f_r$ and $\langle {f_r} \rangle$ is the expectation of $f_{ri}$. Applying the calculus of variations, we write the Lagrangian:
\begin{equation}
L = 
- \sum\limits_{i=1}^I p_i \ln \frac{p_i}{q_i}   
- \lambda_0 \biggl( \sum\limits_{i=1}^I p_i - 1 \biggr) 
- \sum\limits_{r = 1}^R \lambda_r \biggl( \sum\limits_{i = 1}^I {p_i f_{ri} }- \langle {f_r } \rangle \biggr)
\label{eq:Lagr}
\end{equation}
where $\lambda_r$ is the Lagrangian multiplier for the $r$th constraint. Maximising \eqref{eq:Lagr} then gives the most probable realization and maximum relative entropy \cite{Jaynes_1957, Jaynes_1963, Jaynes_2003}:
\begin{empheq}[box=\fbox]{gather}
%\begin{split}
p_i^* =  \frac{q_i}{Z} \exp \Bigl(  - \sum\limits_{r = 1}^R  \lambda_r  f_{ri} \Bigr),
\hspace{15pt} \text{with}  \hspace{5pt}
Z =  e^{\lambda_0} = \sum\limits_{i=1}^I q_i \exp \Bigl(  - \sum\limits_{r = 1}^R  \lambda_r  f_{ri} \Bigr)
%\end{split}
\label{eq:p_i}
\\
\mathfrak{H}^* =  \ln Z  +  \sum\limits_{r = 1}^R \lambda_r \langle f_r \rangle  = -\Phi +  \sum\limits_{r = 1}^R \lambda_r \langle f_r \rangle 
\label{eq:MaxREnt}
\end{empheq}
where $^*$ denotes the inferred state, $Z$ is the partition function and $\Phi = - \ln Z$ is the potential (negative Massieu) function. By further analysis of first and second derivatives under this generic framework, it can be shown that $\mathfrak{H}^*(\langle f_1 \rangle, ..., \langle f_R \rangle)$ and $\Phi(\lambda_1, ..., \lambda_R)$ are Legendre transforms \cite{Jaynes_1957, Jaynes_1963, Jaynes_2003}. 

{A caveat to the foregoing analysis is that the MaxEnt method is not a method of deductive reasoning, but should instead be viewed as a method of probabilistic inference \cite{Jaynes_1957, Jaynes_2003, Niven_2009_PRE, Niven_2010_PTB, Niven_2012_AIP, Dewar_M_B2L}. The distribution inferred by MaxEnt is not necessarily the ``most correct'' representation, but simply the one which is most probable given the imposed choices of constraints, prior probabilities, state space and the relative entropy function itself.  If these assumptions are incomplete or incorrect, the discrepancy will be incorporated in the resulting model.  Furthermore, there may be dynamical restrictions which prevent a system from attaining its most probable state. Such phenomenology (metastable states, supersaturated solutions, reaction kinetics, etc) is well-known in equilibrium thermodynamics and, if necessary, can be handled by the incorporation of additional constraints, restrictions to the state space or additional theoretical apparatus. }

\subsection{\label{sec:S}Thermodynamic Entropy}
\begin{figure}[t]
\begin{center}
\setlength{\unitlength}{0.6pt}
  \begin{picture}(500,145)
   %\put(30,0){(a)}
   \put(50,0){\includegraphics[width=95mm]{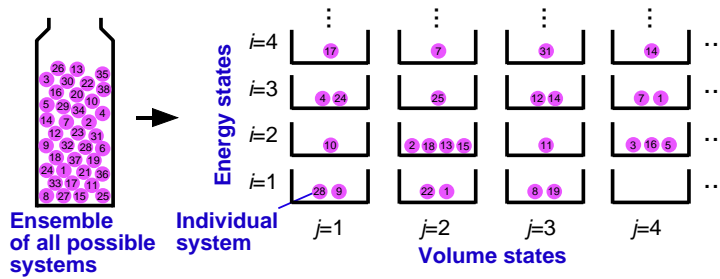}}
   %\put(300,0){(b)}
  \end{picture}
\end{center}
\caption{Allocation scheme for the canonical ensemble of equilibrium thermodynamics.}
\label{fig:canon}
\end{figure}

The thermodynamic entropy $S$ can now be interpreted as a special case of the generic entropy $\mathfrak{H}$, for a physical system constrained by its {\it contents} (usually expressed by mean extensive variables).   
{Consider a container of $N$ interacting molecules, for which it is infeasible to examine the allocation of individual molecules to energetic or other states.  We therefore consider the {\it canonical ensemble} of all possible configurations of the system \cite{Gibbs_1902, Schrodinger_1952, Hill_1956, Atkins_1982, Callen_1985, Niven_2012_CD}, in which replicas of the system are allocated to a coupled bivariate classification scheme according to their energy $\epsilon_{ij}$ and volume $V_{ij}$, where $i$ and $j$ respectively index the discrete energy and volume states of the ensemble.  This is illustrated schematically in Figure \ref{fig:canon}. The probabilities $p_{ij}$ of the $ij$th energy-volume state of the ensemble are then considered to be constrained by normalisation \eqref{eq:Cr}, the mean internal energy $U = \sum\nolimits_{ij} p_{ij} \epsilon_{ij}$ and mean volume $V = \sum\nolimits_{ij} p_{ij} V_{ij}$. Adopting the bivariate relative entropy $\mathfrak{H} =  - \sum\nolimits_{ij} p_{ij} \ln ({p_{ij}}/{q_{ij}})$, the Lagrangian is:}
\begin{equation}
L = - \sum\limits_{ij} p_{ij} \ln \frac{p_{ij}}{q_{ij}}   
- \lambda_0 \biggl( \sum\limits_{ij} p_{ij} - 1 \biggr) 
- \lambda_U \biggl( \sum\limits_{ij} {p_{ij} \epsilon_{ij} }- U \biggr)
- \lambda_V \biggl( \sum\limits_{ij} {p_{ij} V_{ij} } - V \biggr)
\label{eq:Lagr_UV}
\end{equation}
where $\lambda_U$ and $\lambda_V$ are Lagrangian multipliers for $U$ and $V$. 
Maximisation then yields the most probable realization and maximum relative entropy:   
\begin{gather}
%\begin{split}
p_{ij}^* =  \frac{q_{ij}}{Z} \exp \bigl(  - \lambda_U  \epsilon_{ij} - \lambda_V V_{ij} \bigr),
\hspace{15pt} \text{with}  \hspace{5pt}
Z =   \sum\limits_{ij} q_{ij} \exp \bigl(  - \lambda_U  \epsilon_{ij} - \lambda_V V_{ij} \bigr)
%\end{split}
\label{eq:p_i_UV}
\\
\mathfrak{H}^* = \ln Z +  \lambda_U U  +  \lambda_V V = -\Phi +  \lambda_U U  +  \lambda_V V
\label{eq:MaxREnt_UV}
\end{gather}
{These are interpreted to represent the inferred or {\it equilibrium} state of the ensemble \cite{Jaynes_1957}.}
From the empirical body of thermodynamics, or from monotonic considerations, we recognise $\lambda_U = 1/kT$ and $\lambda_V = P/kT$, where $k$ is Boltzmann's constant, $T$ is absolute temperature and $P$ is absolute pressure, while $q_{ij} = \gamma_{ij}/\sum\nolimits_{ij} \gamma_{ij}$ is commonly expressed in terms of the degeneracy $\gamma_{ij}$ of the $ij$th energy-volume level.  Furthermore, we can identify $S = k \mathfrak{H}^*$ as the  thermodynamic entropy at equilibrium, while $\phi_G = k \Phi = G/T$ is the Planck potential\footnote{Strictly, Planck used the negative of $\phi_G$ as his potential function \cite{Planck_1914, Planck_1922}.}, wherein $G$ is the Gibbs free energy. Eqs.\ \eqref{eq:p_i_UV}-\eqref{eq:MaxREnt_UV} thus provide the core equations of equilibrium thermodynamics \cite{Jaynes_1957, Jaynes_1963, Jaynes_2003, Tribus_1961a, Tribus_1961b, Callen_1985}:
\begin{gather}
%\begin{split}
p_{ij}^* =  \frac{\gamma_{ij}}{\hat{Z}} \exp \biggl( \frac{ - \epsilon_{ij} - PV_{ij} }{kT} \biggr),
\hspace{10pt} \text{with}  \hspace{5pt}
\hat{Z} =   {Z}\;{\sum\limits_{ij} \gamma_{ij}} =   \sum\limits_{ij} \gamma_{ij} \exp \biggl( \frac{ - \epsilon_{ij} - PV_{ij} }{kT} \biggr)
%\end{split}
\label{eq:p_i_UV2}
\\
{S} =  k \ln Z +  \frac{U}{T}  +  \frac{PV}{T} = - \phi_G +  \frac{U}{T}  +  \frac{PV}{T}  %= - \phi_G +  \frac{H}{T}  
\label{eq:MaxREnt_UV2}
\end{gather}
%where $H$ is the enthalpy. 
Further analysis using generalised heat and work concepts \cite{Jaynes_1957} gives the differential:
\begin{gather}
%\boxed{
%d{S} =  - d \phi_G + \frac{1}{T} dU +  \frac{P}{T} dV
%}
%\hspace{12pt} \text{or} \hspace{12pt}
\boxed{
d{ \phi_G} =  - dS + \frac{1}{T} dU +  \frac{P}{T} dV 
}
\label{eq:Gibbs_UV}
\end{gather}
Eqs.\ \eqref{eq:p_i_UV2}-\eqref{eq:Gibbs_UV} in turn give a set of derivative relations and Legendre duality between $S$ and $\phi_G$ \cite{Jaynes_1957, Jaynes_1963, Jaynes_2003, Tribus_1961a, Tribus_1961b, Kapur_K_1992, Callen_1985}.  Many other formulations are available for different thermodynamic ensembles subject to various constraints \cite{Tribus_1961b, Kapur_K_1992, Hill_1956}.  

We can now interpret the physical meaning of the potential $\phi_G$  \cite{Niven_2009_PRE, Planck_1922, Callen_1985, Guggenheim_1967}. Consider a ``universe'' divided into a system of interest and an external environment. From the second law \eqref{eq:2ndlaw}, an incremental increase in entropy of the universe can be expressed as a sum of changes within and external to the system $dS_{univ} = dS + dS_{ext} \ge 0$. Although $dS_{ext}$ cannot be measured directly, if it alters the system in any way, it must produce a change in its constraints and/or multipliers, hence $dS_{ext} = - \frac{1}{T} dU -  \frac{P}{T} dV$, where the negative sign accounts for positive $dS_{ext}$. Substituting in \eqref{eq:Gibbs_UV}, we identify $dS_{univ} = - d \phi_G$. In consequence, if a thermodynamic system can interact with an external environment, its equilibrium state is determined by minimising its Planck potential $\phi_G$, thereby maximising the entropy of the universe, rather than by maximising the entropy of the system $S$ alone.  For constant $T$, this reduces to the well-known principle of minimum Gibbs free energy \cite{Gibbs_1875}. 

{
Minimising $\phi_G$ requires integration $\Delta \phi_G  = \int\nolimits_{\path \in \paths} d \phi_G$ over some path $\path$, selected from the set of paths $\paths$ with a specified starting point $\phi_{G,0}$ and an endpoint at the minimum potential $\phi_{G,\min}$. Since $\phi_G$ is a state function, its difference $\Delta \phi_G = \phi_{G,\min} - \phi_{G,0}$ is path-independent, but there may be restrictions on the set of allowable paths $\paths$ (e.g., only adiabatic paths or only isobaric paths), causing further restrictions on the minimum potential $\phi_{G,\min}$, or the set of such minima, which can be accessed by the system.  Denoting $d \sigma = - (dU +  P dV)/T = -d H/T$ as the increment of  {\it entropy produced} by a system, where $H$ is the enthalpy, \eqref{eq:Gibbs_UV} reduces to $d{ \phi_G} =  - dS -d \sigma$. Since $S$ and $\sigma$ are also state functions, the step change can be written as $\Delta{ \phi_G} =  - \Delta S - \Delta \sigma$.  Minimisation of $\phi_G$ to give $\Delta \phi_G<0$ can therefore occur in three ways:
\begin{enumerate}
\item By a coupled increase in both $S$ and $\sigma$ along path $\path$ to give $\Delta \phi_G<0$, hence with $\Delta S>0$ and $\Delta \sigma>0$;
\item By a coupled increase in $S$ and decrease in $\sigma$ along $\path$, hence $\Delta S>0$ and $\Delta \sigma <0$, provided that $\Delta S>| \Delta \sigma |>0$ to ensure $\Delta \phi_G<0$; or
\item By a couple decrease in $S$ and increase in $\sigma$ along $\path$, hence $\Delta S<0$ and $\Delta \sigma >0$, provided that $\Delta \sigma > |\Delta S| >0$ to ensure $\Delta \phi_G<0$.
\end{enumerate}
The choice of scenario is governed by the set of allowable paths $\paths$, which controls the flow of various quantities (in this example, heat) through the control surface and hence the competition between $dS$ and $d\sigma$. The first and third scenarios can be interpreted as a constrained maximisation of $\sigma$ (hence minimisation of $H/T$) over the set of paths $\paths$, while the second can be viewed as a constrained minimisation. Similarly, the first and second scenarios also involve constrained maximisation of $S$ over $\paths$, while the third involves its minimisation. This three-fold structure is well established in equilibrium thermodynamics, although is usually presented in terms of the Gibbs free energy rather than the Planck potential \cite{Atkins_1982}. Rather than adopt separate extremum principles for different processes, and to correctly account for changes in entropy within and outside the system, the three scenarios are unified by an overarching minimum Planck potential principle \cite{Planck_1922, Guggenheim_1967}, which at constant $T$ reduces, as noted, to that of minimum Gibbs free energy \cite{Gibbs_1875}. }

{
As will be shown, the above thermodynamically-inspired principle can be established -- using the MaxEnt framework -- in other, quite different kinds of systems.}

\subsection{\label{sec:EB}Entropy Balance and Entropy Production}

\textbf{Entropy Balance Equations:} 
With the entropy concept in hand, we can now consider the thermodynamic entropy balance in a control volume, such as that shown in Figure \ref{fig:CV}(a). Our first difficulty is that $S$ is not conserved. However, from the second law of thermodynamics, within any closed physical system:
\begin{equation}
dS \ge 0  %\hspace{20pt} \text{or} \hspace{20pt} \frac{DS_{FV(t)}}{Dt} \ge 0
\label{eq:2ndlaw}
\end{equation}
%*** excursions??**
where $dS$ implies a mean differential over a minimum time scale, to allow for brief excursions in the opposite sense.  So, despite not being conserved, we can say that $S$ is {\it preserved}: once created, it cannot be destroyed. In consequence, for an entropically open system -- which can exchange entropy with its external environment -- \eqref{eq:Bbalance} provides a control volume balance (``{\it law of preservation}'') for $S$:
\begin{equation}
\frac{DS_{FV(t)}}{Dt} = \frac{\partial S_{CV}}{\partial t} + \F_{S,f}^{out}  - \F_{S,f}^{in}  
\label{eq:Sbalance1} 
\end{equation}
where $\F_{S,f}^{out}$ and $\F_{S,f}^{in}$ are the outflow and inflow rates of $S$ due to fluid flow through the control surface. The substantial derivative can also be separated, by the de Donder technique, into externally- and internally-driven rates of change of entropy within the fluid volume, giving the overall entropy balance equation (c.f. \cite{Moran_S_2006}):
\begin{equation}
\boxed{
\frac{DS_{FV(t)}}{Dt} 
= \frac{\partial S_{CV}}{\partial t} + \F_{S,f}^{out} - \F_{S,f}^{in}  
=  \frac{D_{e}S_{FV(t)}}{Dt} + \frac{D_{i}S_{FV(t)}}{Dt} 
}
\label{eq:Sbalance2} 
\end{equation}
where ${D_{e}S_{FV(t)}}/{Dt}$ represents the rate of change of entropy in the fluid volume due to non-fluid flows (positive inwards), i.e.
\begin{equation}
\frac{D_{e}S_{FV(t)}}{Dt}  = \F_{S,nf}^{in} - \F_{S,nf}^{out}
\end{equation}
Similarly, ${D_i S_{FV(t)}}/{Dt}$ denotes the {\it (rate of) entropy production} in the fluid volume due to internal processes, henceforth labelled $\dot{\sigma}$. The latter serves as a book-keeping term in \eqref{eq:Sbalance2}, ensuring that the rate of creation of entropy in the fluid volume satisfies the second law of thermodynamics \eqref{eq:2ndlaw}:
\begin{equation}
\boxed{
\dot{\sigma} = \frac{D_{i}S_{FV(t)}}{Dt} 
%= \frac{\partial S_{CV}}{\partial t} + \F_{S,f}^{out} - \F_{S,f}^{in}  - \frac{D_{e}S_{FV(t)}}{Dt} 
%= \frac{\partial S_{CV}}{\partial t} + \F_{S,f}^{out} + \F_{S,nf}^{out} - \F_{S,f}^{in}  - \F_{S,nf}^{in} 
= \frac{\partial S_{CV}}{\partial t} + \F_{S,tot}^{out} - \F_{S,tot}^{in} 
\ge 0 
}
\label{eq:EPdef} 
\end{equation}
where $\F_{S,tot}=\F_{S,f}+\F_{S,nf}$ is the total entropy flow rate. 
Thus by definition, the rate of entropy production $\dot{\sigma}$ cannot be negative, regardless of whether the newly created entropy is retained in the control volume or exported from it (i.e., {independent of the sign of the rate of change of $S$}).  Eq.\ \eqref{eq:EPdef} may therefore be viewed as a powerful manifestation of the second law, applicable to all non-equilibrium systems.

For the integral control volume of Figure \ref{fig:CV}(b), from \eqref{eq:Re_transp}:%-\eqref{eq:Re_transp_div}:
\begin{equation}
\frac{DS_{FV(t)}}{Dt} 
= \frac{\partial }{\partial t} \iiint\limits_{CV} \rho s dV  + \oiint\limits_{CS} \rho s \vec{v} \cdot \vec{n} dA 
%= \iiint\limits_{CV} \biggl[  \frac{\partial }{\partial t} \rho s  +   \nabla \cdot \rho s \vec{v} \biggr] dV
\label{eq:Re_transp_S} 
\end{equation}
where $s$ is the specific entropy. From \eqref{eq:Sbalance2}, this is equal to the internal rate of entropy production in the fluid volume, $\dot{\sigma}$, plus the external rate of input due to non-fluid transport processes, $- \oiint\nolimits_{FS(t)} \vec{j}_S \cdot \vec{n} dA$, where $\vec{j}_{S}$ is the non-fluid entropy flux:
\begin{equation}
\frac{DS_{FV(t)}}{Dt} 
= \frac{D_{i}S_{FV(t)}}{Dt} + \frac{D_{e}S_{FV(t)}}{Dt}
= \dot{\sigma} - \oiint\limits_{FS(t)} \vec{j}_S \cdot \vec{n} dA
\label{eq:Re_transp_S2} 
\end{equation}
Equating \eqref{eq:Re_transp_S}-\eqref{eq:Re_transp_S2}, for coincident fluid and control volumes at time $t$, gives:
\begin{equation}
\boxed{
\dot{\sigma} 
= \iiint\limits_{CV}\frac{\partial \rho s}{\partial t} dV  + \oiint\limits_{\substack{
\text{coincident} \\ 
CS \text{ and } FS(t)}}
\bigl[ \vec{j}_S + \rho s \vec{v} \bigr] \cdot \vec{n} dA
= \frac{DS_{FV(t)}}{Dt}  + \oiint\limits_{FS(t)} \vec{j}_S \cdot \vec{n} dA
}
%\cdot \vec{n} dA  + \oiint\limits_{FS(t)}
\label{eq:Re_transp_S3} 
\end{equation}
Applying \eqref{eq:Re_transp_mass} and Gauss' theorem then yields:
\begin{equation}
%\boxed{
\dot{\sigma} 
= \iiint\limits_{CV} \biggl[  \frac{\partial }{\partial t} \rho s  + \nabla \cdot \vec{J}_{S} \biggr ] dV   
= \iiint\limits_{FV(t)} \biggl[ \rho \frac{Ds}{Dt}  + \nabla \cdot \vec{j}_S \biggr] dV
%}
\label{eq:Re_transp_S4} 
\end{equation}
where $\vec{J}_{S} = \vec{j}_{S} + \rho s \vec{v}$. Finally, subdividing $\dot{\sigma}  = \iiint\nolimits_{CV} \hat{\dot{\sigma}} dV$, where $\hat{\dot{\sigma}}$ %= \partial (\rho s)^{int} / \partial t$ 
is the {\it (rate of) entropy production per unit volume},
and equating integrands (assuming validity at all scales) gives the differential entropy balance equation \cite{Kondepudi_P_1998, deGroot_M_1984, Bird_etal_2006, Jaumann_1911}:
\begin{equation}
\boxed{
\hat{\dot{\sigma}} 
=   \frac{\partial }{\partial t} \rho s + \nabla \cdot \vec{J}_{S} 
=   \rho \frac{Ds}{Dt}   +  \nabla \cdot \vec{j}_S
%=   \frac{\partial }{\partial t} \rho s + \nabla \cdot (\rho s \vec{v}) + \nabla \cdot \vec{J}_{S} 
\ge 0
}
\label{eq:cons_diff_S} 
\end{equation}
By a scale invariance argument \cite{Kondepudi_P_1998}, $\hat{\dot{\sigma}}$ cannot {be negative} locally (at least over a minimum time scale) at any location, since this would continuously destroy thermodynamic entropy {within an identifiable control volume}, and so violate the second law of thermodynamics. This is entirely separate to the {rate of change of the specific entropy $s$, which can be positive or negative locally}, depending on the sign of the divergence term (i.e. on the local entropy flux out of the element).  

\textbf{Local Entropy Flux and Entropy Production:} %
To reduce \eqref{eq:cons_diff_S}, we seek functional forms of the non-fluid entropy flux $\vec{j}_S$ and local entropy production $\hat{\dot{\sigma}}$. For non-radiative processes, the standard approach is to start from the substantial derivative of the specific form of Gibbs' relation \eqref{eq:MaxREnt_UV2}-\eqref{eq:Gibbs_UV} \cite{Prigogine_1967, deGroot_M_1984, Bird_etal_2006, Jaumann_1911}:
\begin{equation}
\frac{Ds}{Dt} = \frac{D \phi_g}{Dt}  +   \frac{1}{T} \frac{D u}{Dt}  +   \frac{P}{T} \frac{D \rho^{-1}}{Dt} 
\label{eq:D_Gibbs_UV}
\end{equation}
where $\phi_g = g/T$ is the specific Planck potential and $g$ the specific Gibbs free energy. This adopts the {\it local equilibrium assumption}, where each infinitesimal element is assumed to be in thermodynamic equilibrium and so can be described by local intensive variables $1/T$, $P/T$ and $\{\mu_c/T\}$, where $\mu_c$ is the molar chemical potential of species $c$. Including the work of chemical diffusion $g = - \sum\nolimits_c \mu_c n_c$, and substituting for the substantial derivatives of specific volume $\rho^{-1}$, species molar densities $n_c$ and specific internal energy $u$ (see Table \ref{tab:cons}) gives:
\begin{equation}
\rho \frac{Ds}{Dt} = 
 - \frac{1}{T} \nabla \cdot \vec{j}_Q 
 + \sum\limits_{c} \frac{\mu_c}{T} \, \nabla \cdot \vec{j}_c 
% + \frac{1}{T} \,  \sum\limits_c M_c \vec{j}_c \cdot \vec{g}_c 
 - \frac{1}{T} \,  \sum\limits_c M_c \vec{j}_c \cdot \nabla \psi_c 
 -  \frac{1}{T} \, \tens{\tau} : \nabla \vec{v}
 -  \sum\limits_{d} \hat{\dot{\xi}}_{d} \, \Delta \frac{\tilde{G}_{d}}{T}
\label{eq:D_Gibbs_UV2}
\end{equation}
This is expressed in terms of the molar rate of the $d$th reaction $\hat{\dot{\xi}}_d = \sum\nolimits_c \chi_{cd} \hat{\dot{\xi}}_c$ ($>0$ if a product) and change in molar Planck potential of the $d$th reaction, $\Delta \tilde{\phi}_G = \Delta (\tilde{G}_{d}/T)=\sum\nolimits_{c} \chi_{cd} \, {\mu}_c/T$ ($<0$ if spontaneous), where $\chi_{cd}$ is the stoichiometric coefficient of species $c$ in the $d$th reaction.  Comparison to \eqref{eq:cons_diff_S}, with some vector calculus, gives the entropy flux and local entropy production \cite{deGroot_M_1984, Bird_etal_2006}:
\begin{empheq}[box=\fbox]{gather}
{\vect{j}}_{S,m} = \biggl( {\frac{1}{T}} \biggr) {\vect j}_Q  
- \sum\limits_{c} \biggl( \frac{\mu_c}{T} \biggr) {\vect j}_c  
%{ - \sum\limits_{c} {\vect j}_c \biggl( \frac{{\psi}_c}{T} \biggr)  -  {{\tens{\tau}}} \cdot  \biggl( \frac{ { \vect v} }{T} \biggr)} 
%{ - \sum\limits_{c} {\vect j}_c z_{c} \biggl( \frac{{\psi}_z}{T} \biggr) **}
\label{eq:j_s}
\\
%\begin{split}
\hat{\dot{\sigma}}_m= 
{\vect j}_Q \cdot \nabla \biggl( {\frac{1}{T}} \biggr) 
%- \sum\limits_{c} {{\vect j}_c} \cdot \biggl[ \nabla \biggl( \frac{\mu_c}{T} \biggr) - \frac {M_c {\vect g}_c}{T}   \biggr]
- \sum\limits_{c} {{\vect j}_c} \cdot \biggl[ \nabla \biggl( \frac{\mu_c}{T} \biggr) + \frac {M_c \nabla \psi_c}{T}   \biggr]
%\\
 -  \frac{\tens{\tau} : \nabla \vec{v} }{T}
 -  \sum\limits_{d} \hat{\dot{\xi}}_{d} \, \Delta \frac{\tilde{G}_{d}}{T}
% {**charge}
%\end{split}
\label{eq:sigma_dot_hat2}
\end{empheq}
These do not include the effect of radiation, examined in a later section, and so are labelled $m$ to signify the material or thermodynamic component. In generic form, we identify the entropy flux \eqref{eq:j_s} as $\vec{j}_{S,m} = \sum\nolimits_{r} \vec{j}_{r} \lambda_{r}$, a sum of products of fluxes and conjugate spatial intensive variables selected from $\vec{j}_r \in \{ \vec{j}_Q, \vec{j}_c \}$ and $\lambda_r \in \{1/T,$ $ -\mu_c/T \}$, %, $ $-\psi_c/T, $ $-\vec{v}/T \}$, 
while the entropy production \eqref{eq:sigma_dot_hat2} is $\hat{\dot{\sigma}}_m=  \sum\nolimits_{r} \vec{j}_{r} \cdot \vec{F}_{r}$, a sum of products of all fluxes or rates and their conjugate gradients or driving forces $\vec{F}_r \in \{\nabla (1/T),$ $- \nabla (\mu_c/T), $ $-\nabla \psi_c/T, $ $-\nabla \vec{v}/T, $ $- \Delta (\tilde{G}_d/T) \}$ \cite{Prigogine_1967, Kondepudi_P_1998, deGroot_M_1984}. %\footnote{$\Delta (\tilde{G}_d /T)$ can be considered a gradient in chemical composition space, rather than physical space.}. 
Usually, $\hat{\dot{\sigma}}_m$ is further simplified -- assuming conditions close to thermodynamic equilibrium -- using the linear Onsager phenomenological relations and the Curie postulate, to give a bilinear sum of thermodynamic forces \cite{deGroot_M_1984, Bird_etal_2006, Kjelstrup_etal_2010}. 

%{**Comment: I have gone through this analysis many times, following several authors' approaches. I verify the equations, but cannot understand why there is no momentum term $- (\vec{v} /T)  \cdot \tens{\tau}$ within $\vec{j}_S$. Does it vanish for a silly reason, e.g. tensor symmetry?  The body force term also vanishes, but I suspect this is assumed to be a reversible work divided by T.}

%
\begin{figure}[t]
\begin{center}
\setlength{\unitlength}{0.6pt}
  \begin{picture}(500,175)
   \put(30,0){(a)}
   \put(60,0){\includegraphics[width=85mm]{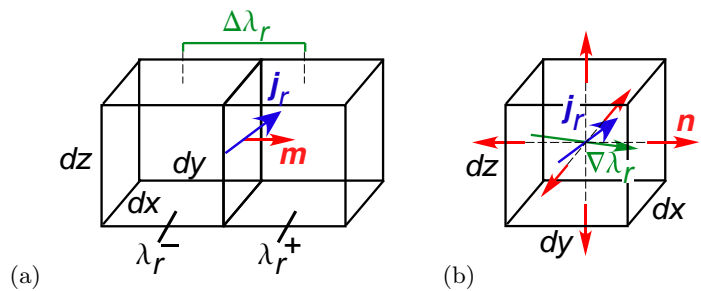}}
   \put(300,0){(b)}
    \end{picture}
\end{center}
\caption{Infinitesimal volume elements for (a) local spatial equilibrium (Type I and II(a)) and (b) continuum (Type II(b)) representations, showing the $r$th flux and its intensive variables.}
\label{fig:vol_el}
\end{figure}

\textbf{Thermodynamic Representations:} Before embarking on further analyses, it is worth scrutinising the physical representation of the bilinear, non-radiative local entropy production \eqref{eq:sigma_dot_hat2}. As evident, it includes two quite different types of physical processes:
\newcounter{Lcount1}
\begin{list}{{\textbf {\textit {Type \Roman{Lcount1} Processes:}}}}{\usecounter{Lcount1} \topsep 2pt \itemsep 2pt \parsep 0pt \leftmargin 0pt \rightmargin 0pt \listparindent 0pt \itemindent 17pt}
\item Those which can be represented to occur within an  infinitesimal volume element at {\it local spatial equilibrium} with respect to the spatial intensive variables $\lambda_r \in \{1/T, -\mu_c/T, -\psi_c/T, -\vec{v}/T \}$, as shown in Fig. \ref{fig:vol_el}(a)\footnote{Some authors unite the variables conjugate to the species flux $\vec{j}_c$ into a local electrochemical or gravichemical potential divided by temperature, $-\mu_c^{\vec{g}}/T$ \cite{Guggenheim_1967}.}. In \eqref{eq:sigma_dot_hat2}, only the final chemical reaction term falls into this category.  In this case, the volume element need not be in chemical equilibrium, but may be maintained at a higher Planck potential by its chemical composition. This category also includes nuclear and subatomic decay processes, not usually represented in \eqref{eq:sigma_dot_hat2}. 
\item Those which -- although formulated in terms of an infinitesimal volume element -- are in fact associated with a physical {\it flux} which diminishes (or acts conjugate to) a spatial {\it gradient}. For the $r$th process, this can be written as $\hat{\dot{\sigma}}_{m,r}=  \vec{j}_r \cdot \nabla \lambda_r$.  The heat, species mass and momentum transport terms in \eqref{eq:sigma_dot_hat2} all fall into this category. These have two possible physical representations:
\end{list}
\newcounter{Lcount2}
\begin{list}{(\alph{Lcount2})}{\usecounter{Lcount2} \topsep 0pt \itemsep 2pt \parsep 0pt \leftmargin 16pt \rightmargin 0pt \listparindent 0pt \itemindent 0pt}
\item If each volume element is considered to be in local spatial equilibrium, as shown in Fig. \ref{fig:vol_el}(a), then no Type II entropy production could occur {\it within} an element, but only {\it between} elements. This necessitates analysis of the boundary entropy production terms, which must be integrated over the internal boundaries and/or somehow assigned to each element.
\item If each volume element need not be in local spatial equilibrium, it can be used to directly represent both the fluxes and gradients, as shown in Fig. \ref{fig:vol_el}(b). Arguably, this gives a more physically defensible representation of a non-equilibrium system -- dependent upon the continuum assumption -- and for this reason is almost universally adopted throughout fluid mechanics and heat transfer analysis (e.g. in differential derivations of the continuity, Navier-Stokes and energy equations). However, it contradicts the assumption of local equilibrium, creating a philosophical difficulty in the use of intensive variables which, strictly, are defined only at equilibrium \cite{Callen_1985}. Instead, in this representation, both a value and gradient in each intensive variable are assigned to each point within the infinitesimal element.
%\item An intermediate representation, consistent with both local equilibrium and volumetric entropy production, involves the use of subdivided volume elements, each subdomain of which is considered in local equilibrium.   
\end{list}
Representations II(a) and II(b) involve fundamentally different idealisations of physical transport processes. Their analysis requires different mathematical tools, respectively a hybrid difference-differential calculus and the usual differential calculus. 

To tease out the distinction between Type II(a)-(b) representations, consider an individual boundary between two infinitesimal elements, as shown in Fig.\ \ref{fig:vol_el}(a)-(b). For Type II(b) elements, there is no discontinuity in the $r$th intensive variable at each boundary, and -- by continuity -- no change in the $r$th flux, causing no (or an infinitesimal) change in each non-fluid entropy flux. The fluid-borne entropy flux  $\rho s \vec{v}$ is similarly unaffected.  In consequence, no (or an infinitesimally small) entropy production occurs at the boundary. Type II(a) elements, in contrast, exhibit a step change $\Delta \lambda_r =  \lambda_r^{+}  -  \lambda_r^{-}$ in each spatial intensive variable across the boundary, giving the {\it net entropy production per unit area} (J K$^{-1}$ m$^{-2}$ s$^{-1}$) due to thermodynamic processes at the boundary: 
\begin{gather}
\EParea_m
= \Bigl[\Delta (\rho s) \vec{v}  + \sum_r {\vec j}_{r} \, \Delta \lambda_{r} \Bigr] \cdot {\vec{m}} 
= \bigl[\Delta (\rho s) \vec{v}  + \Delta {\vec j}_{S,m} \bigr] \cdot {\vec{m}} 
= \Delta {\vec J}_{S,m} \cdot {\vec{m}}
\ge 0 
\label{eq:net_sigma_dot_breve}
\end{gather}
where $\vec{j}_{r}$ is the flux of the $r$th conserved quantity, ${\vec{m}}$ is the unit normal to the boundary, $\Delta {\vec j}_{S,m}$ is the net non-fluid, non-radiative entropy flux and $\Delta {\vec J}_{S,m}$ also includes the net fluid-borne entropy flux (all positive in the direction $\Delta \lambda_r >0$). In \eqref{eq:net_sigma_dot_breve}, it is assumed the fluid-borne entropy flux undergoes a step change at the boundary (e.g. due to a discontinuity $\Delta \rho$ or $\Delta s$ caused by a step change in $1/T$ or $\mu_c/T$). No step changes are considered in $\vec{v}$ or $\vec{j}_r$, being fluxes of conserved quantities. The non-radiative entropy production along a boundary $\Gamma$ is then $\dot{\sigma}_{\Gamma,m} = \iint\nolimits_{\Gamma} \EParea_m \, dA$.  

Often it is desirable to account separately for each side of the boundary, leading to the {\it absolute} or {\it half-boundary entropy production per unit area} due to outward flow from a specified face of a volume element:
\begin{gather}
\halfEParea_m
= \Bigl[ \rho s \vec{v} + \sum_r {\vec j}_{r} \, \lambda_{r} \Bigr] \cdot {\vec{n}} 
= \bigl[ \rho s \vec{v} + {\vec j}_{S,m} \bigr] \cdot {\vec{n}} 
= {\vec J}_{S,m} \cdot {\vec{n}}
\gtrless 0
\label{eq:abs_sigma_dot_breve}
\end{gather}
where $\vec{n}$ is the outward unit normal. As expected, this depends on the material entropy flux $\vec{J}_{S,m}$ at the boundary. From \eqref{eq:net_sigma_dot_breve}, $\EParea_m = \halfEParea_m^{+} - \halfEParea_m^{-}$. The total entropy production along $\Gamma$ is thus given by the two-sided surface integral $\dot{\sigma}_{\Gamma,m}  
= \iint\nolimits_{\Gamma^{+}} \halfEParea_m^{+} \, dA - \iint\nolimits_{\Gamma^{-}} \halfEParea_m^{-} \, dA  
= \oiint\nolimits_{\Gamma} \halfEParea_m \, dA
= \oiint\nolimits_{\Gamma} {\vec J}_{S,m} \cdot {\vec{n}} \, dA$. Applying Gauss' divergence theorem to the surface $\Gamma$ enclosing the ``internal volume'' $\Gamma^o$, we obtain the interesting result that $\dot{\sigma}_{\Gamma,m} = \iiint\nolimits_{\Gamma^o} \nabla \cdot {\vec J}_{S,m} \, dV \ge 0$, even though $\iiint_{\Gamma^o} dV=0$.

From the second law \eqref{eq:2ndlaw}, each net boundary entropy production \eqref{eq:net_sigma_dot_breve} is non-negative (over a minimum observation time). In contrast, the half-boundary terms \eqref{eq:abs_sigma_dot_breve} can be of arbitrary sign, so long as their difference across each internal boundary is non-negative\footnote{In this respect, the half-boundary entropy production terms $\halfEParea_m$ are analogous to half-reaction electrode potentials.}. As a test of consistency, integration of \eqref{eq:abs_sigma_dot_breve} over the external control surface yields the net entropy flow rate $\protect{\oiint\nolimits_{CS} {\vec J}_{S,m} \cdot {\vec{n}} \, dA}$ contained in \eqref{eq:Re_transp_S3}.  %As noted, in general the latter term can be positive or negative, depending on the sign of the integrated rate of change $\iiint\nolimits_{CV} \partial (\rho s)/\partial t \, dV$, so long as the overall entropy production is non-negative. 

Eqs.\ \eqref{eq:net_sigma_dot_breve}-\eqref{eq:abs_sigma_dot_breve} are used in later sections. They cannot, however, be reconciled in a straightforward manner to the differential equation in \eqref{eq:cons_diff_S}, which corresponds strictly to the Type II(b) or continuum representation.

%The analysis of Type II entropy production is therefore quite different to that of other conserved quantities, due to its bilinear form (flux $\times$ gradient). Both conditions must be fulfilled: a flux without a gradient, or a gradient without a flux, do not produce entropy and so must be correctly accounted.

\textbf{Effect of Radiation:} %
An important category of processes, omitted from the standard analysis \eqref{eq:D_Gibbs_UV2}-\eqref{eq:sigma_dot_hat2} -- and indeed from most references on non-equilibrium thermodynamics -- is the entropy production associated with electromagnetic radiation.  Its major principles were however enunciated by Planck \cite{Planck_1901, Planck_1914} over a century ago, and further developed over the past century (e.g.\ \cite{Rosen_1954, Ore_1955, Kroll_1967, Essex_1984a, Essex_1984b, Essex_1987, Callies_Herbert_1988, Pelkowski_1994, Goody_Abdou_1996}).  However, there still remains widespread confusion in its calculation, over choices of symbols and preferred parameters, and even in the most appropriate theoretical approach\footnote{For consistency with this chapter, some notational changes are also necessary here.}. Many renowned texts on radiation omit the topic entirely (e.g.\ \cite{Goody_Yung_1989}). 

Firstly, the energy of unpolarised electromagnetic radiation per unit frequency travelling through an infinitesimal area (of unit normal $\vec{m}$) and infinitesimal solid angle per unit time is represented by its {\it specific energy intensity} or {\it energy radiance} $I_{\nu}$ (SI units: W m$^{-2}$ s sr$^{-1}$). This is a function of the direction $\vec{m}$.  The {\it radiative energy flux} or {\it energy irradiance} (W m$^{-2}$) of radiation striking an infinitesimal area with unit normal $\vec{n}$ is then obtained by integration over all incident directions and the spectrum \cite{Planck_1914}:
\begin{equation}
\vec{j}_{E,\nu} = \vec{n}  \; \int\limits_{0}^{\infty}  \iint\limits_{\Omega(\vec{m})}   I_{\nu} (\vec{m})  \; \vec{m} \cdot \vec{n}  \;  d \Omega (\vec{m})  \, d\nu 
\label{eq:rad_energy_flux}
\end{equation}
where $\Omega$ is the solid angle (in steradians) and $\nu$ is the frequency. Most authors employ $\vec{m} \cdot \vec{n} = \cos \theta$ in \eqref{eq:rad_energy_flux}, with $\theta$ a function of $\vec{m}$.  Here, \eqref{eq:rad_energy_flux} is integrated over a sphere ${\Omega(\vec{m})} \in [0, 4\pi]$ to account for travelling radiation from all directions (the {\it net flux}); for  radiation incident on a solid surface, \eqref{eq:rad_energy_flux} is integrated over a hemisphere ${\Omega(\vec{m})} \in [0, 2\pi]$ (the {\it absolute flux}).  For polarised radiation, the two orthogonal components must be examined separately \cite{Callies_Herbert_1988}; {an even more general description invokes the two-dimensional complex polarisation tensor, involving conservation of linear and angular momentum as well as energy \cite{Landau_Lifshitz_1972}. Note that \eqref{eq:rad_energy_flux} describes a {\it reversible} energy flux; this only becomes irreversible in the event of changes in radiance, which necessarily require the interaction of radiation and matter \cite{Callies_Herbert_1988}. } 

Similarly, we can consider the {\it specific entropy intensity} or {\it entropy radiance} $L_{\nu}$ (W K$^{-1}$ m$^{-2}$ s sr$^{-1}$) of radiation. This is given by
 \cite{Planck_1914, Rosen_1954, Ore_1955, Essex_1984a, Essex_1984b, Essex_1987, Callies_Herbert_1988, Pelkowski_1994}:
\begin{equation}
L_\nu (\vec{m}) = \frac{2 k \nu^2}{{c_0}^2} \biggl[ \biggl(\frac{{c_0}^2 I_\nu (\vec{m})}{2 \h \nu^3} +1 \biggr) \ln  \biggl(\frac{{c_0}^2 I_\nu (\vec{m})}{2 \h \nu^3} +1 \biggr) - \frac{{c_0}^2 I_\nu (\vec{m})}{2 \h \nu^3} \ln  \frac{{c_0}^2 I_\nu (\vec{m})}{2 \h \nu^3}  \biggr]
\label{eq:entropy_radiance}
\end{equation}
where $k$ is Boltzmann's constant, $c_0$ is the speed of light in a vacuum and $\h$ is Planck's constant.  Eq.\ \eqref{eq:entropy_radiance} can be obtained from the Bose-Einstein entropy function, needed to describe electromagnetic radiation \cite{Planck_1901, Planck_1914, Rosen_1954, Ore_1955, Niven_2005, Niven_2006, Niven_2009}, and is a property of the radiation itself, independent of the entropy produced by its conversion to heat.  A different (Fermi-Dirac) relation applies to neutrinos \cite{Essex_Kennedy_1999}. The {\it radiative entropy flux} or {\it entropy irradiance} (W K$^{-1}$ m$^{-2}$) is then given by:
\begin{equation}
\vec{j}_{S,\nu} = \vec{n}  \; \int\limits_{0}^{\infty}  \iint\limits_{\Omega(\vec{m})}   L_{\nu} (\vec{m})  \; \vec{m} \cdot \vec{n}  \;   d \Omega (\vec{m}) \, d\nu
\label{eq:rad_entropy_flux}
\end{equation}
For unpolarised radiation emitted from a black-body of temperature $T$, the specific energy intensity is given by the well-known Planck equation \cite{Planck_1901, Planck_1914}:
\begin{equation}
I_\nu = 2 B_\nu =  \frac{2 \hbar \nu^3}{{c_0}^2} \frac{1}{\exp (\hbar \nu /kT) -1}  
\label{eq:Planck_law}
\end{equation}
whereupon \eqref{eq:rad_entropy_flux} reduces to $|\vec{j}_{S,\nu}| = \frac{4}{3} k_{SB} T^3$, where $k_{SB}$ is the Stefan-Boltzmann constant \cite{Essex_1984a, Essex_1984b}. 

We can now construct the local entropy production as the sum of non-radiative (material) and radiative components \cite{Essex_1984a, Essex_1984b, Essex_1987, Callies_Herbert_1988, Pelkowski_1994, Goody_Abdou_1996}:
\begin{align}
\hat{\dot{\sigma}} = \hat{\dot{\sigma}}_m + \hat{\dot{\sigma}}_\nu
\label{eq:EP_withrad1}
\end{align}
From \eqref{eq:cons_diff_S}, applicable equally to either component:
\begin{align}
\hat{\dot{\sigma}}_m
&=   \frac{\partial }{\partial t} \rho s + \nabla \cdot \vec{J}_{S,m} 
=   \frac{\partial }{\partial t} \rho s + \nabla \cdot \vec{j}_{S,m} + \nabla \cdot (\rho s \vec{v})
\label{eq:EP_mat}
\\
\hat{\dot{\sigma}}_\nu 
&=   \frac{\partial }{\partial t} \hat{S}_\nu + \nabla \cdot \vec{j}_{S,\nu}
\label{eq:EP_rad}
\end{align}
where $\hat{S}_\nu$ is the entropy per volume due to radiation. 
Note that only the radiative entropy flux $\vec{j}_{S,\nu}$ appears in \eqref{eq:EP_rad}; the Clausius heating term $\vec{j}_{E,\nu}/T$ due to the radiative energy flux must be incorporated into the thermodynamic entropy flux in \eqref{eq:EP_mat} \cite{Essex_1984a, Essex_1984b}. Putting these together, the total local entropy production due to material processes and radiation is:
\begin{align}
\boxed{
\hat{\dot{\sigma}}
=   \frac{\partial }{\partial t} \rho s 
+  \frac{\partial }{\partial t} \hat{S}_\nu 
+ \nabla \cdot (\rho s \vec{v})
+ \nabla \cdot \vec{j}_{S,m}  + \nabla \cdot \vec{j}_{S,\nu}
}
\label{eq:EP_withrad2}
\end{align}

To reduce \eqref{eq:EP_withrad2}, several approaches have been taken in the literature. Essex \cite{Essex_1984a, Essex_1984b} applies a volumetric form of the Gibbs equation \eqref{eq:MaxREnt_UV2} and total energy conservation, for flows only of heat, radiation and chemical constituents, to give:    
\begin{align}
%\boxed{
\hat{\dot{\sigma}}
= \biggl\{ {\vect j}_Q \cdot \nabla \biggl( {\frac{1}{T}} \biggr) 
- \sum\limits_{c} {{\vect j}_c} \cdot \nabla \biggl( \frac{\mu_c}{T} \biggr) 
\biggr\}
  - \frac{1}{T} \frac{\partial }{\partial t} \hat{U}_\nu +  \frac{\partial }{\partial t} \hat{S}_\nu
%+ \rho s \vec{v}
- \frac{ \nabla \cdot \vec{j}_{E,\nu}}{T}  
+ \nabla \cdot \vec{j}_{S,\nu}
%}
\label{eq:EP_withrad_Essex}
\end{align}
where $\hat{U}_\nu$ is the energy per volume of radiation, and the braces enclose the material component. Essex \cite{Essex_1987} extended this to fluid flows with viscous dissipation. Alternatively, Callies \& Herbert \cite{Callies_Herbert_1988} and Goody \& Abdou \cite{Goody_Abdou_1996} adopt a Gibbs-like equation for radiation:
\begin{align}
d\hat{S}_\nu= \frac{1}{T_\nu} d \hat{U}_\nu
\label{eq:Gibbs_rad}
\end{align}
where $T_\nu$ is a radiative temperature, defined based on \eqref{eq:Planck_law} as the temperature of matter in equilibrium with radiation of frequency $\nu$. Non-black-body radiation can thus exhibit different radiative temperatures at different wavelengths. 
For heat and radiative transport only, this leads to \cite{Kroll_1967, Essex_1984b, Callies_Herbert_1988, Goody_Abdou_1996}:
\begin{align}
%\boxed{
\hat{\dot{\sigma}}
= {\vect j}_Q \cdot \nabla \biggl( {\frac{1}{T}} \biggr) 
+ \int\limits_{0}^{\infty}    \iint\limits_{\Omega(\vec{m})}  
%\frac{d I_\nu (\vec{m})}{dr}  
\biggl( \frac{1}{c_0} \frac{\partial I_\nu}{\partial t} + \vect{m} \cdot \nabla I_\nu \biggr)
\; \biggl( \frac{1}{T_\nu (\vec{m})} - \frac{1}{T} \biggr)  \;  d \Omega (\vec{m})  \, d\nu  
%}
\label{eq:EP_withrad_CH}
\end{align}
%where ${d I_\nu}/{dr}$ is the derivative of energy radiance in its direction of propagation. 
{Kr\" oll \cite{Kroll_1967} and} Callies \& Herbert  \cite{Callies_Herbert_1988} argue that the integral in \eqref{eq:EP_withrad_CH} provides a bilinear formulation of the radiative entropy production, with the first term in brackets (the source function) behaving as an extensive variable.  Essex \cite{Essex_1984a, Essex_1984b}, however, disputes this view, since the bilinearity applies to each wavelength and direction. In any case, further corrections are needed in the event of scattering. 

As pointed out by Essex \cite{Essex_1984b}, integration of the local radiative entropy production \eqref{eq:EP_withrad2} over a control volume is not straightforward, due to the emission and absorption of radiation {by non-adjacent volume elements}.  This creates direct, {non-local} connections between every element $dV$, creating a very different control volume to those usually {examined} in fluid mechanics. This gives the entropy production term:
\begin{align}
{\dot{\sigma}}_\nu^{heat} = \frac{1}{2} \iiint\limits_{CV} \iiint\limits_{CV} h(\vec{x}_1, \vec{x}_2) \biggl( \frac{1}{T(\vec{x}_2)} - \frac{1}{T(\vec{x}_1)} \biggr) dV dV
\label{eq:nonlocal1}
\end{align}
where $h(\vec{x}_1, \vec{x}_2) $ %\nabla \cdot \vec{j}_{E,\nu} /dV
is the net rate at which heat from position vector $\vec{x}_1$ is delivered to $\vec{x}_2$ via radiation. Allowing for the loss of energy and entropy radiation from the control volume then gives:
\begin{align}
{\dot{\sigma}}_\nu 
= \frac{1}{2} \iiint\limits_{CV} \iiint\limits_{CV} h(\vec{x}_1, \vec{x}_2) \biggl( \frac{1}{T(\vec{x}_2)} 
- \frac{1}{T(\vec{x}_1)} \biggr) dV dV 
- \iiint\limits_{CV} \frac{f_{CS}(\vec{x})}{T(\vec{x})} dV 
+ \iint\limits_{CS} \vec{j}_{S,\nu} \cdot \vec{n} \, dA
\label{eq:nonlocal2}
\end{align}
where $f_{CS}(\vec{x})$ is the component of radiative energy from position $\vec{x}$ which escapes through the control surface. For a control volume which completely encloses a planet, all terms in \eqref{eq:nonlocal2} and all material terms vanish except the entropy radiation, giving $\dot{\sigma} = \iint\nolimits_{CS}\, \vec{j}_{S,\nu} \cdot \vec{n} \, dA$ \cite{Essex_1984a, Callies_Herbert_1988}. On these grounds, Essex \cite{Essex_1984a} argues against the MaxEP hypothesis of Paltridge \cite{Paltridge_1975}, on the grounds that the dominant, radiative entropy production term is missing. 

{A rather different approach for radiative transfer, involving a minimum entropy production closure of the radiative energy flux \eqref{eq:rad_energy_flux} and higher-order moments, is outlined in \cite{Christen_K_B2L}.} Further treatments of entropy production due to radiative absorption, scattering and other interactions lie beyond the scope of this chapter, and are discussed in the above-cited works.

\textbf{Compartmentalisation:} %
For many applications, it is desirable to subdivide a control volume into $K$ contiguous compartments. From $\dot{\sigma}  = \iiint\nolimits_{CV} \hat{\dot{\sigma}} dV$, it might be assumed that the global entropy production is simply the sum of that in each compartment. However, this depends on the representation used. For compartments composed of Type II(b) elements, with no intensive variable discontinuities at their boundaries, this assumption is correct. If, however, the compartments are composed of Type II(a) elements, it is also necessary to account for the entropy production due to flows {\it between} compartments.  In consequence, for purely material flows:
\begin{align}
\begin{split}
{\dot{\sigma}}_m 
&= \sum\limits_{\alpha=1}^K \dot{\sigma}_m^{\alpha}
+   \sum\limits_{\alpha=2}^K \sum\limits_{\beta=1}^{\alpha-1} \; \iint\limits_{CS_{\alpha \beta}} \EParea_m^{\alpha \beta} \, dA
\\
&= \sum\limits_{\alpha=1}^K \dot{\sigma}_m^{\alpha}
+   \sum\limits_{\alpha=2}^K \sum\limits_{\beta=1}^{\alpha-1} \; \iint\limits_{CS_{\alpha \beta}} 
\Bigl[ \Delta (\rho s)^{\alpha \beta} \vec{v}^{\alpha \beta} + \sum_r {\vec j}_{r}^{\alpha \beta} \, \Delta \lambda_{r}^{\alpha \beta} \Bigr] \cdot {\vec{m}} \, dA
\end{split}
\label{eq:EP_incl_bound}
\end{align}
where ${\dot{\sigma}}_m^{\alpha}$ is the material entropy production in the $\alpha$th compartment, while $\EParea^{\alpha \beta}_m$ is the material entropy production per area on the control surface $CS_{\alpha \beta}$ between the $\alpha$th and $\beta$th compartments (counted only once and for $\alpha \ne \beta$). In terms of bulk flow rates:
%For the special case in which each compartment is in spatial but not necessarily chemical equilibrium (Type II processes) -- a common assumption in planetary atmospheric models -- the first term in \eqref{eq:EP_incl_bound} will contain only chemical reaction or Type I processes, while the second term may simplify to a sum of bulk net entropy flow rates:
\begin{equation}
{\dot{\sigma}}_m 
= \sum\limits_{\alpha=1}^K \dot{\sigma}^{\alpha}_m
+   \sum\limits_{\alpha=2}^K \sum\limits_{\beta=1}^{\alpha-1} \; \biggl[ \Delta \F_{S,f}^{{\alpha \beta}} + \Delta \F_{S,nf,m}^{{\alpha \beta}}  \biggr] 
%+   \sum\limits_{\alpha=2}^K \sum\limits_{\beta=1}^{\alpha-1} \; \biggl[\Delta \F_{S,f}^{{\alpha \beta}} + \sum_r \F_{r}^{nf,{\alpha \beta}} \, \Delta \lambda_{r}^{\alpha \beta} \biggr] 
= \sum\limits_{\alpha=1}^K \dot{\sigma}^{\alpha}_m
+   \sum\limits_{\alpha=2}^K \sum\limits_{\beta=1}^{\alpha-1} \; \Delta  \F_{S,m}^{{\alpha \beta}} 
\label{eq:bulk_EP_incl_bound}
\end{equation}
where $\Delta \F_{S,f}^{ \alpha \beta}$, $\Delta \F_{S,nf,m}^{{\alpha \beta}}$ and $\Delta \F_{S,m}^{{\alpha \beta}}$ respectively designate the bulk net fluid-borne, non-fluid (non-radiative) and total thermodynamic entropy flow rates normal to the $\alpha \beta$ control surface. If radiative transfer can also take place, \eqref{eq:EP_incl_bound}-\eqref{eq:bulk_EP_incl_bound} must be augmented by the three terms in \eqref{eq:nonlocal2}, with attention to boundary transitions.  Relations \eqref{eq:EP_incl_bound}-\eqref{eq:bulk_EP_incl_bound} do not require steady state; by definition \eqref{eq:EPdef}, each measurable entropy production term is independently non-negative and therefore additive.

%{In the presence of radiation, *** Essex nonlocality**  FINISH}

\textbf{Steady State:} %
Since most entropy-producing systems involve fluctuating conditions,  a strict steady state \eqref{eq:Re_transp_stst} or \eqref{eq:cons_diff_stst} is not meaningful. We thus consider the mean steady state ${\partial \langle S \rangle _{CV}}{\partial t}=0$, for which the bulk balance \eqref{eq:Sbalance2} gives, in general:
\begin{align} 
\boxed{
\langle \dot{\sigma} \rangle
= \sum\limits_{\kappa \in CS} \Bigl [ \langle \F_{S,f}^{\kappa} \rangle  + \langle \F_{S,nf}^{\kappa} \rangle \Bigr ]
=\sum\limits_{\kappa \in CS}  \langle \F_{S, tot}^{\kappa} \rangle
}
\label{eq:Sbalance_st1} 
\end{align}
where $\langle \F_{S,f}^{ \kappa} \rangle$, $\langle \F_{S,nf}^{\kappa} \rangle$ and $\langle \F_{S,tot}^{\kappa} \rangle$ are respectively the bulk mean fluid-borne, non-fluid and total entropy flow rates through portion $\kappa$ of the control surface. Similarly, \eqref{eq:Re_transp_S3}-\eqref{eq:Re_transp_S4} and \eqref{eq:cons_diff_S} (hence \eqref{eq:EP_withrad2}) give, respectively:
\begin{empheq}[box=\fbox]{align}
%\begin{align}
\frac{\partial \langle S \rangle_{CV}}{\partial t}=0  
\hspace{10pt} &\Rightarrow \hspace{10pt} 
\langle \dot{\sigma} \rangle 
= \oiint\limits_{CS} \langle \vec{J}_{S} \rangle \cdot \vec{n} dA 
= \iiint\limits_{CV}  \nabla \cdot \langle \vec{J}_{S} \rangle dV
\label{eq:EPmean_global} 
\\
\frac{\partial }{\partial t} \langle \rho s \rangle = 0 
\hspace{10pt} &\Rightarrow \hspace{10pt} 
\langle \hat{\dot{\sigma}} \rangle
=    \nabla \cdot \langle  \vec{J}_{S} \rangle 
=   \nabla \cdot \langle  \rho s \vec{v} \rangle + \nabla \cdot \langle  \vec{j}_S \rangle 
\label{eq:EPmean_local} 
\end{empheq}
%\end{align}
%
From \eqref{eq:Sbalance_st1}-\eqref{eq:EPmean_local}, the mean steady state is quite special, since under this condition, all of the entropy production is exported from the control volume. This restricts the total mean entropy flow terms in \eqref{eq:Sbalance_st1}-\eqref{eq:EPmean_local} to be nonnegative.  Accordingly, at mean steady state, the total mean entropy production can be calculated either by integration of the mean of \eqref{eq:sigma_dot_hat2} over the control volume, or more directly from the sum \eqref{eq:Sbalance_st1} or integral \eqref{eq:EPmean_global} of mean entropy flows through the control surface. 

%Returning to the analysis of bulk compartments, each in spatial (but not necessarily chemical) equilibrium, we see from \eqref{eq:bulk_EP_incl_bound}-\eqref{eq:Sbalance_st1} that for material flows at steady state:
%\begin{equation}
%\boxed{
%\langle \dot{\sigma}_m \rangle
%= \sum\limits_{\alpha=1}^K \langle \dot{\sigma}^{\alpha}_I \rangle
%+   \sum\limits_{\alpha=2}^K \sum\limits_{\beta=1}^{\alpha-1} \; \langle \Delta \F_{S,m}^{{\alpha \beta}} \rangle  
%=\sum\limits_{\kappa \in CS}  \langle \F_{S, m}^{\kappa} \rangle
%}
%\label{eq:bulk_EP_intextbal_stst}
%\end{equation}
%or in the presence of radiation:
%\begin{equation}
%%****FINISH
%\label{eq:bulk_EP_intextbal_stst_withrad}
%\end{equation}

We therefore see that \eqref{eq:Sbalance_st1}-\eqref{eq:EPmean_global} express an {\it internal-external entropy balance}: at mean steady state, the {\it total} mean entropy produced within a control volume %in the compartments and at sharp internal boundaries 
will exactly balance the {\it total} mean entropy flow out of its external boundaries.  Often this is assumed without proof, but it requires the mean steady state, and applies only to the total {quantities}. In the presence of radiation, the radiative transport terms must be included {within these totals}. 

\subsection{\label{sec:closure} Reynolds-Averaged Entropy Production and Closure Problem} %
%{**CAN I REVISE IN LIGHT OF RADIATION**}

We now raise an objection to one feature of previous studies of the MaxEP principle or hypothesis, as applied to planetary climate and other fluid flow systems \cite{Paltridge_1975, Paltridge_1978, Ozawa_etal_2003, Kleidon_L_book_2005, Martyushev_S_2006, Bruers_2007c}. This objection applies only to the material (non-radiative) component of time-varying, stationary flows, amenable to the Reynolds decomposition and averaging method \cite{Spurk_1997, Durst_2008, Schlichting_2001, White_2006}. Although not stated explicitly, the vast majority of such studies do not actually use the mean steady-state entropy production $\langle \hat{\dot{\sigma}} _m \rangle =  \sum\nolimits_{\ell} \langle \vec{j}_{\ell} \cdot \vec{F}_{\ell} \rangle$ \eqref{eq:EPmean_local} or its global form \eqref{eq:Sbalance_st1}-\eqref{eq:EPmean_global}.  Instead, they invoke a different quantity: the {\it steady-state entropy production in the mean}, $\lmean \hat{\dot{\sigma}}_m \rmean =  \sum\nolimits_{\ell} \langle \vec{j}_{\ell} \rangle \cdot \langle \vec{F}_{\ell} \rangle$, based on products of mean fluxes or rates and their conjugate mean gradients or forces.  These two quantities are not the same. 
By Reynolds decomposition of each independent quantity $a = \langle a \rangle + a'$, where $a'(\vec{x},t)$ is the time-varying component, subject to the usual averaging rules\footnote{Typical Reynolds averaging rules for irreducible parameters $a$ and $b$ are:
$\langle 1 \rangle = 1$, 
$\langle \langle a \rangle \rangle = \langle a \rangle$, 
$\langle a + b \rangle = \langle a \rangle + \langle b \rangle$,
$\langle a \langle b \rangle \rangle = \langle a \rangle \langle b \rangle$,
$\langle a' \rangle =0$, 
%$\langle a' \langle b \rangle \rangle =0$, 
$\langle \partial a/\partial x \rangle = \partial \langle a \rangle /\partial x$ and 
$\langle \int a dx \rangle = \int \langle a \rangle dx$
\cite{Spurk_1997, Durst_2008, Schlichting_2001, White_2006}.}, 
the difference is:
%whence:
\begin{equation}
\begin{split}
\lf \hat{\dot{\sigma}}_m \rf 
=&
\langle \hat{\dot{\sigma}}_m \rangle - \lmean \hat{\dot{\sigma}}_m \rmean 
=  \sum\nolimits_\ell \langle \vec{j}_{\ell}' \cdot \vec{F}_{\ell}' \rangle
\\
=& 
\biggl\langle {\vect j}_Q' \cdot \nabla \biggl( {\frac{1}{T}} \biggr)'  \biggr\rangle
- \sum\limits_{c} \biggl\langle {{\vect j}_c'} \cdot \nabla \biggl( \frac{\mu_c}{T} \biggr)' \biggr\rangle
% &+ \sum\limits_{c} \biggl\langle {{\vect j}_c'} \cdot  \biggl( \frac {M_c {\vect g}_c}{T} \biggr)' \biggr\rangle
- \sum\limits_{c} \biggl\langle {{\vect j}_c'} \cdot  \biggl( \frac {M_c \nabla \psi_c}{T} \biggr)' \biggr\rangle
\\ &
-  \biggl\langle \tens{\tau}' :  \biggl( \frac{ \nabla \vec{v}}{T} \biggr)' \biggr\rangle
 -  \sum\limits_{d} \biggl\langle \hat{\dot{\xi}}_{d}' \, \Delta \biggl( \frac{\tilde{G}_{d}}{T} \biggr) ' \biggr\rangle 
 \ge 0
\end{split}
\label{eq:sigma_diff}
\end{equation}
Usually the flux and rate terms in \eqref{eq:sigma_diff} are linearised using Onsager coefficients as functions of the forces, giving a sum of quadratic fluctuation terms (see \cite{Kock_Herwig_2004, Naterer_Camberos_2008}). Depending on its cause, the body force may be strictly steady and so disappear from \eqref{eq:sigma_diff}. All other terms, however, consist of nonzero nonlinear products, except under strict steady-state conditions. %, or in peculiar systems with uncoupled fluxes and gradients. 

In dissipative systems far from equilibrium, the mean fluctuating entropy production $\lf \hat{\dot{\sigma}}_m \rf $ \eqref{eq:sigma_diff} can be considerably larger -- in many cases by orders of magnitude -- than the entropy production in the mean $\lmean \hat{\dot{\sigma}}_m \rmean$ \cite{Spurk_1997, Durst_2008, Schlichting_2001, White_2006}. It is therefore difficult, {\it a priori}, to see why the latter should constitute the objective function for a variational principle. As shown in \S\ref{sec:FlowSys}, however, precisely this function emerges from a judicious MaxEnt analysis of a non-equilibrium system at steady state.

We now incorporate fluctuating radiation with mean entropy production $\langle \hat{\dot{\sigma}}_\nu \rangle$ and mean net entropy flux $\langle \vec{j}_{S,\nu} \rangle$. 
%and mean net fluctuations $\langle \vec{j}_{S,\nu}' \rangle=0$
Writing $\lmean \vec{j}_{S,m} \rmean = \sum\nolimits_{\ell} \langle \vec{j}_{\ell} \rangle \langle \lambda_{\ell} \rangle$ for the material entropy flux in the mean and $\lf \vec{j}_{S,m} \rf  = \sum\nolimits_{\ell} \langle \vec{j}_{\ell}' \lambda_{\ell}' \rangle$ for its mean fluctuation, Reynolds averaging of the local entropy balance \eqref{eq:EP_withrad2} yields:
\begin{equation}
\boxed{
\begin{split}
\langle  \hat{\dot{\sigma}} \rangle =
\lmean \hat{\dot{\sigma}}_m \rmean 
+ \lf  \hat{\dot{\sigma}}_m \rf 
+ \langle  \hat{\dot{\sigma}}_\nu \rangle 
%\\
= \nabla \cdot 
\Bigl\{
\lmean \rho s \vec{v} \rmean
+ \lf \vec{\varsigma}_{\vec{v}}  \rf
+ \lmean  \vec{j}_{S,m} \rmean     
+ \lf  \vec{j}_{S,m} \rf
+ \langle \vec{j}_{S,\nu} \rangle
\Bigl\}
%\\
%\text{with} \hspace{5pt}
\end{split}
}
\label{eq:cons_diff_S_decomp}
\end{equation}
%with \vspace{-14pt}
with $\lmean \rho s \vec{v} \rmean = \langle  \rho \rangle \langle s \rangle \langle \vec{v} \rangle$ and  
$\lf \vec{\varsigma}_{\vec{v}}  \rf  =  
\langle \rho ' s' \rangle \langle \vec{v} \rangle 
+ \langle s' \vec{v}' \rangle \langle \rho \rangle 
+ \langle \rho' \vec{v}' \rangle \langle s \rangle 
+ \langle \rho ' s' \vec{v} ' \rangle$. 
On integration and application of Gauss' theorem:
\begin{equation}
\begin{split}
%\boxed{
\langle  {\dot{\sigma}} \rangle 
&= \iiint\limits_{CV} \Bigl\{ 
\lmean \hat{\dot{\sigma}}_m \rmean
+ \lf \hat{\dot{\sigma}}_m \rf 
+ \langle  \hat{\dot{\sigma}}_\nu \rangle 
\Bigr\} dV
%= \iiint\limits_{CV} \lmean \hat{\dot{\sigma}}_m \rmean dV
%+ \iiint\limits_{CV} \lf \hat{\dot{\sigma}}_m \rf dV
\\
&=
\oiint\limits_{CS} \Bigl\{ 
\lmean \rho s \vec{v} \rmean 
+ \lmean  \vec{j}_{S,m} \rmean  
\Bigr\} \cdot \vec{n} dA
+ \oiint\limits_{CS} \Bigl\{ 
\lf \vec{\varsigma}_{\vec{v}}  \rf 
+ \lf  \vec{j}_{S,m} \rf  \Bigr\}  
\cdot \vec{n} dA
+ \oiint\limits_{CS} 
 \langle \vec{j}_{S,\nu} \rangle
\cdot \vec{n} dA
%}
\end{split}
\label{eq:Re_transp_S_decomp} 
\end{equation}
or, in macroscopic terms:
\begin{align} 
\boxed{
\langle  {\dot{\sigma}} \rangle =
\lmean \dot{\sigma}_m \rmean 
+ \lf \dot{\sigma}_m \rf
+ \langle  {\dot{\sigma}}_\nu \rangle 
=\sum\limits_{\kappa \in CS}  \Bigl \{ 
\lmean \F_{S, m}^{\kappa} \rmean  
+ \lf {\F_{S,m}^{\kappa}} \rf  
+ \langle {\F_{S,\nu}^{\kappa}} \rangle  \Bigr \}
}
\label{eq:Sbalance_st1_decomp} 
\end{align}
The non-vanishing mean fluctuation terms of the material flows in \eqref{eq:cons_diff_S_decomp}-\eqref{eq:Sbalance_st1_decomp} create many difficulties.  Firstly, there is no guarantee -- even at steady state -- that the material entropy production in the mean $\lmean \dot{\sigma}_m \rmean$ will be in balance with the net outward material entropy flow in the mean $\sum\nolimits_{\kappa \in CS}  \lmean \F_{S,m}^{ \kappa} \rmean $. In other words, it is possible that part of the mean fluctuating component of the material entropy production $\lf \dot{\sigma}_m \rf$ is converted into outward entropy flows in the mean $\lmean \F_{S,m}^{ \kappa} \rmean$, or into the mean radiative flux $\langle {\F_{S,\nu}^{\kappa}} \rangle$. Alternatively, some of the material entropy production in the mean $\lmean \dot{\sigma}_m \rmean$ could be converted into mean fluctuating entropy flows $\lf {\F_{S,m}^{\kappa}} \rf$ or carried by radiation.  It is therefore not possible to claim, without further proof, that the extremum calculated using one of $\lmean \dot{\sigma}_m \rmean$  or $\sum\nolimits_{\kappa \in CS}  \lmean \F_{S,m}^{\kappa} \rmean $, or one such term plus its corresponding radiative term, is equivalent to the extremum based on the other. %A similar difficulty applies to the macroscopic internal-external total entropy balance \eqref{eq:bulk_EP_intextbal_stst}, in which the quantities in the mean need not balance. 
Secondly, it is not possible, even in principle, to calculate the fluctuation terms from the mean quantities, since they contain additional unknown (and correlated) parameters, unless some other theoretical principles or constitutive relations can be invoked.  

These features of fluctuating, dissipative flow systems are well-known in fluid mechanics, but are here generalised to all non-equilibrium systems with fluid and non-fluid flows.  They can collectively be referred to as the {\it entropy production closure problem}.
%\footnote{Some fluid mechanicists would refer to this as a second closure problem, the first being the requirement to adopt the continuum hypothesis \cite{**}.}. 
This problem affects the vast majority of previous studies on entropy production extremum principles, in which the distinction between in-the-mean and total mean components is not taken explicitly into account. 

%{Would it help to outline a simple example here, e.g. of a heat flow system?}

\section{\label{sec:FlowSys}MaxEnt Analysis of Flow Systems}

%{**MUST REVISE IN LIGHT OF RADIATION**}
We now close our discussion of control volume analysis and entropy balance by a direct MaxEnt analysis of a flow system \cite{Niven_2009_PRE, Niven_2010_PTB, Niven_2012_AIP}. This provides a fundamental framework for the analysis of non-equilibrium systems -- indeed, as fundamental as thermodynamics itself -- yet underpinned by the same generic foundation provided by Jaynes' method. The analysis can be applied at any scale, integral or differential \cite{Niven_2012_AIP}; here we only examine the local scale, in the absence of radiation.

Consider an infinitesimal volume element within a control volume, as shown in Figure \ref{fig:CV}(b), using the Type II(b) continuum representation. Such a fluid element experiences instantaneous values of various fluxes and rates $\vec{j}_{\ell, \vec{i}} \in \{ \vec{j}_Q, \vec{j}_c, \tens{\tau}, \hat{\dot{\xi}}_d \}$. At the mean steady state, these are constrained by their mean values $\langle \vec{j}_{\ell}  \rangle \in \{ \langle \vec{j}_Q \rangle, \langle \vec{j}_c \rangle,$ $\langle \tens{\tau} \rangle, \langle \hat{\dot{\xi}}_d \rangle \}$. We therefore adopt the multivariate relative entropy $\mathfrak{H}_{st} =  - \sum\nolimits_{\vec{i}} p_{\vec{i}} \ln ({p_{\vec{i}}}/{q_{\vec{i}}})$ -- here termed the {\it flux  entropy} \cite{Niven_2009_PRE} -- as a measure of the variability or uncertainty in the allocation of fluxes and rates to possible instantaneous values. Combining the entropy and constraints gives the Lagrangian:
\begin{equation}
L_{st} = - \sum\limits_{\vec{i}} p_{\vec{i}} \ln \frac{p_{\vec{i}}}{q_{\vec{i}}}   
- \zeta_0 \Bigl( \sum\limits_{\vec{i}} p_{\vec{i}} - 1 \Bigr) 
- \sum\limits_{\ell} \vec{\zeta}_{\ell} \cdot \Bigl( \sum\limits_{\vec{i}} p_{\vec{i}} \, \vec{j}_{\ell, \vec{i}} - \langle \vec{j}_{\ell} \rangle \Bigr)
\label{eq:Lagr_flow}
\end{equation}
where $\zeta_0$ and $\zeta_\ell$ are Lagrangian multipliers for  normalisation and the $\ell$th constraint. 
Maximisation yields the most probable realization and maximum flux entropy:   
\begin{gather}
%\begin{split}
p_{\vec{i}}^* =  \frac{q_{\vec{i}}}{Z_{st}} \exp \Bigl(  - \sum\limits_\ell \vec{\zeta}_\ell \cdot \vec{j}_{\ell, \vec{i}}   \Bigr),
\hspace{15pt} \text{with}  \hspace{5pt}
Z_{st} =   \sum\limits_{\vec{i}} q_{\vec{i}} \exp \Bigl(  - \sum\limits_\ell \vec{\zeta}_\ell \cdot \vec{j}_{\ell, \vec{i}}   \Bigr)
%\end{split}
\label{eq:p_i_flow}
\\
\mathfrak{H}_{st}^* 
= \ln Z_{st} +  \sum\limits_\ell \vec{\zeta}_\ell \cdot \langle \vec{j}_{\ell} \rangle  
= - \Phi_{st} +  \sum\limits_\ell \vec{\zeta}_\ell \cdot \langle \vec{j}_{\ell} \rangle \label{eq:MaxREnt_flow}
\end{gather}
where $Z_{st}$ is the flux partition function and $\Phi_{st} = - \zeta_0$ can be interpreted as a local {\it flux potential} for non-equilbrium systems, analogous to the Planck potential in equilibrium thermodynamics.  Comparing \eqref{eq:MaxREnt_flow}
to the local material entropy production \eqref{eq:sigma_dot_hat2}, we recognise the multipliers as proportional to the mean gradients or forces:
\begin{gather}
\vec{\zeta}_\ell = - \frac{ \langle \vec{F}_\ell \rangle}{\mathcal{K}}
\in 
\frac{1}{\K}
\biggl \{- 
\biggl\langle  \vec{\nabla}  \frac{1}{T} \biggr\rangle, 
\biggl\langle  \vec{\nabla}  \frac{\mu_c}{T} 
+ \frac {M_c \nabla \psi_c}{T}  \biggr\rangle, 
 \biggl\langle \frac{\vec{\nabla}  \vec{v}^{\top} }{T} \biggr\rangle, 
\biggl\langle \Delta \frac{\tilde{G}_{d}}{T} \biggr\rangle
\biggr \}
\label{eq:zeta_rels}
\end{gather}
where $\K$ is a {positive} constant (J K$^{-1}$ m$^{-3}$ s$^{-1}$). {Eqs \eqref{eq:p_i_flow}-\eqref{eq:MaxREnt_flow} then give:}
\begin{gather}
p_{\vec{i}}^* =  \frac{q_{\vec{i}}}{Z_{st}} \exp \frac {\hat{\dot{\sigma}}_{m,\vec{i}}}{\K}
\label{eq:p_i_flow2}
\\
\mathfrak{H}_{st}^* = - \Phi_{st} -  \frac{\lmean \hat{\dot{\sigma}}_m \rmean}{\K}  
\label{eq:MaxREnt_flow2}
\end{gather}
where ${\hat{\dot{\sigma}}_{m,\vec{i}}} =  \sum\nolimits_\ell \langle \vec{F}_\ell \rangle \cdot \vec{j}_{\ell, \vec{i}} $ is the local material entropy production for the $\vec{i}$th category or state, based on mean gradients or forces. We therefore obtain a Gibbs-like relation \eqref{eq:MaxREnt_flow2} for a steady-state flow system, analogous to \eqref{eq:MaxREnt_UV2} for equilibrium systems, which contains the local material entropy production in the mean ${\lmean \hat{\dot{\sigma}}_m \rmean}$.  Further analyses, analogous to those in \S\ref{sec:genericH}, provide a set of derivative relations and Legendre duality between $\mathfrak{H}_{st}^*$ and $\Phi_{st}$ \cite{Niven_2009_PRE, Niven_2010_PTB, Niven_2012_AIP}.   

%Incidentally, since negative flux states $\vec{i}$ are permitted, \eqref{eq:p_i_flow2} gives a generic form of the Fluctuation Theorem \cite{**}:
%\begin{gather}
%\frac{p_{\vec{i}}^*}{p_{-\vec{i}}^*} 
%=  \frac{q_{\vec{i}}}{q_{-\vec{i}}} \exp \frac{ {\hat{\dot{\sigma}}_{\vec{i}}} - {\hat{\dot{\sigma}}_{-\vec{i}}} }{\K}
%=  \frac{q_{\vec{i}}}{q_{-\vec{i}}} \exp   \sum\limits_\ell \frac {\langle \vec{F}_\ell \rangle \cdot (\vec{j}_{\ell, \vec{i}}  -  \vec{j}_{\ell, -\vec{i}})} {\K}  
%\label{eq:FT}
%\end{gather}

{
Just as in equilibrium thermodynamics (see \S\ref{sec:S}), we can interpret the potential $\Phi_{st}$ as the state function which is {\it minimised} to give the most probable state of a ``universe'', consisting of the flow system (control volume) and its controlling environment. Rewriting \eqref{eq:MaxREnt_flow2} using generalised heat and work concepts \cite{Jaynes_1957}:
\begin{gather}
\boxed{
d \Phi_{st}  = - d \mathfrak{H}_{st}^* - \frac{d \lmean \hat{\dot{\sigma}}_m \rmean}{\K}  
}
\label{eq:Phi_st}
\end{gather}
in which each quantity $\Phi_{st}$, $\mathfrak{H}_{st}^*$ and $\lmean \hat{\dot{\sigma}}_m \rmean$ is a state function, we again obtain the step change $\Delta \Phi_{st}  = - \Delta \mathfrak{H}_{st}^* - { \Delta \lmean \hat{\dot{\sigma}}_m \rmean}/{\K}$, given by integration $\int\nolimits_{\path_{st} \in \paths_{st}} \Delta \phi_{st}$ over some path $\path_{st}$ from a set of allowable paths $\paths_{st}$.  We again see that minimisation of $\Phi_{st}$ to give $\Delta \Phi_{st}<0$ can occur in three ways:
\begin{enumerate}
\item By a coupled increase in both $\mathfrak{H}_{st}^{*}$ and ${\lmean \hat{\dot{\sigma}}_m \rmean}$ along $\path_{st}$, whence $\Delta \mathfrak{H}_{st}^{*}>0$ and $\Delta {\lmean \hat{\dot{\sigma}}_m \rmean}>0$;
\item By a coupled increase in $\mathfrak{H}_{st}^{*}$ and decrease in ${\lmean \hat{\dot{\sigma}}_m \rmean}$ along $\path_{st}$, such that $\Delta \mathfrak{H}_{st}^{*}>| \Delta {\lmean \hat{\dot{\sigma}}_m \rmean}/\K |>0$; or
\item By a coupled decrease in $\mathfrak{H}_{st}^{*}$ and increase in ${\lmean \hat{\dot{\sigma}}_m \rmean}$ along $\path_{st}$, such that $\Delta {\lmean \hat{\dot{\sigma}}_m \rmean}/\K > |\Delta \mathfrak{H}_{st}^{*}| >0$.
\end{enumerate}
The first and third scenarios can be interpreted as a constrained maximisation of the entropy production (MaxEP) in the mean, over the set of paths $\paths_{st}$. In contrast, the second scenario can be viewed as a constrained minimisation of the entropy production (MinEP) in the mean, over $\paths_{st}$. Such interpretations do not, however, represent the whole picture, since they fail to  account for changes in the flux entropy $\mathfrak{H}_{st}^*$, which can also be interpreted as being maximised in scenarios 1 and 2 and minimised in scenario 3.  For maximum generality, the three scenarios can be united into a {\it minimum flux potential} principle which controls the state of an infinitesimal flow system. 
}

Further treatments of this analysis are available elsewhere \cite{Niven_2009_PRE, Niven_2010_PTB, Niven_2012_AIP, Noack_Niven_2012, Noack_Niven_2013}. An integral formulation can also be developed, applicable to an entire control volume at mean steady state \cite{Niven_2012_AIP}. The connection between global and local formulations -- especially a formulation which includes radiation \eqref{eq:EP_withrad2} or which takes account of the entropy production closure problem (\S\ref{sec:closure}) -- remains unresolved and requires further research.

To summarise, the foregoing MaxEnt analysis indicates that there is no universal MaxEP or MinEP principle applicable to non-equilibrium flow systems.  Instead, such ``principles'' emerge -- in the mean -- as subsidiary effects under particular conditions. This conclusion is supported by convincing experimental evidence, at least at the integral scale of analysis. This includes inversion of the Paltridge MaxEP principle for fluid flow in pipes, subject either to constraints on the flow rates or the conjugate pressure gradients \cite{Thomas_1942, PaulusJr_G_2004, Martyushev_2007, Niven_2010_JNET}. {Analogous extremum inversions are also observed or suggested by theoretical analyses of plasmas \cite{Kawazura_Yoshida_2010, Kawazura_Yoshida_2012}, turbulent shear flows \cite{Ozawa_etal_2001}, Rayleigh-B\' enard convection \cite{Weaver_etal_B2L}, heat or momentum transfer with advection \cite{Ozawa_S_B2L} and flows around particles \cite{Vaidya_B2L}.} According to the present analysis, such phenomena will be governed by a more general principle involving minimisation of some quantity, related to the flux potential $\Phi_{st}$; the ongoing challenge is to enlarge the underlying theoretical framework.

\section{\label{sec:Concl}Conclusions}

{
This chapter explores the foundations of the entropy and entropy production concepts, using the engineering tool of ``control volume analysis'' for the analysis of fluid flow systems.  Firstly, the principles of control volume analysis are enunciated and applied to flows of conserved quantities (e.g. mass, momentum, energy) through a control volume, giving integral (Reynolds transport theorem) and differential forms of the conservation equations. Strict (instantaneous) and mean definitions of the steady state are provided, based on a stationary first moment or ``Reynolds average''.  The generic entropy concept $\mathfrak{H}$ -- and the purpose of the maximum entropy (MaxEnt) principle -- are established by combinatorial arguments (the Boltzmann principle). An entropic analysis of an equilibrium thermodynamic system is then conducted, giving the thermodynamic entropy $S$. Control volume analyses of a flow system then gives the ``entropy production'' concept for simple, integral and infinitesimal flow systems. Some technical features of such systems are then examined, including discrete and continuum representations of volume elements, the effect of radiation, and the analysis of systems subdivided into compartments. A Reynolds decomposition of the entropy production equation then reveals an ``entropy production closure problem'' in fluctuating dissipative systems: even at steady state, the entropy production based on mean flow rates and gradients is not necessarily in balance with the outward entropy fluxes based on mean quantities.  Finally, the direct application of Jaynes' MaxEnt method yields a theoretical framework with which to predict the steady state of a flow system. This is cast in terms of a ``minimum flux potential'' principle, which reduces, in different circumstances, to maximum or minimum entropy production (MaxEP or MinEP) principles based on mean flows and gradients.  
}

{
Further, substantial research is required on many of the formulations presented in this chapter, especially on the newly disclosed entropy production closure problem (\S\ref{sec:closure}) and on the MaxEnt analysis of steady-state flow systems (\S\ref{sec:FlowSys}). Within the MaxEnt formulation, the effects of local to global scaling (see \S\ref{sec:CV} and \cite{Niven_2009_PRE, Niven_2010_PTB, Niven_2012_AIP, Noack_Niven_2012}) and compartmentalisation (\S\ref{sec:EB}); of time versus ensemble averaging and associated ergodic and transient effects; of non-local interactions by electromagnetic, neutrino or other radiation (\S\ref{sec:EB}); and of the closure problem (\S\ref{sec:closure}), remain unresolved.
It is hoped that this chapter inspires others to attain a deeper understanding and higher technical rigour in the calculation and extremisation of the entropy production in flow systems of all types.
}

\section*{Acknowledgments}

The authors acknowledge funding from the Chair of Excellence ‘Closed-loop control of turbulent shear flows using reduced-order models’ (TUCOROM) of the French Agence Nationale de la Recherche (ANR), hosted by Institute PPrime, Poitiers, France; and a Rector Funded Visiting Fellowship and travel funding from UNSW. We appreciate valuable stimulating discussions at the MaxEP workshop, Canberra, Sept. 2011 and at MaxEnt 2012, Garching, Germany, July 2012, and with Filip Meysman, Bjarne Andresen and Markus Abel. {The three reviewers are sincerely thanked for their comments.}

\pagebreak

\section{\label{sec:Notation}Notation}
{
\begin{longtable}{p{40pt} p{380pt}}
\hline\noalign{\smallskip}
\textbf{Symbol}&\textbf{Meaning (SI Units)}\\
\hline\noalign{\smallskip}
\endfirsthead
\hline\noalign{\smallskip}
\textbf{Symbol}&\textbf{Meaning (SI Units)}\\
\hline\noalign{\smallskip}
\endhead
\hline\noalign{\smallskip}
\endfoot
\hline\noalign{\smallskip}
\endlastfoot
\multicolumn{2}{l}{\textbf{Roman symbols} }\\
$A$ &area (m$^2$)\\
%$b$ &specific (per fluid mass) density of $B$ ($[B]$ kg$^{-1}$)\\
$B$, $b$ &conserved quantity ($[B]$); specific (per fluid mass) density ($[B]$ kg$^{-1}$) \\
$c_0$ &speed of light in vacuum (m s$^{-1}$) \\
$f$, $\vec{F}$ &generic parameter; generic gradient or driving force (various)\\
$\F_B$ &bulk flow rate of quantity $B$ ($[B]$ s$^{-1}$)\\
${\vec g}_c$ &specific body force on species $c$ (N kg$^{-1}$ = m s$^{-2}$)\\
$G$, $g$ &Gibbs free energy (J); specific Gibbs free energy (J kg$^{-1}$)\\
$\Delta \tilde{G}_{d}$ &change in molar Gibbs free energy of reaction $d$ (J  mol$^{-1}$)\\
%$H$ &enthalpy (J) \\
$h$ &net heat transfer rate by radiation (J s$^{-1}$ m$^{-6}$) \\
$\mathfrak{H}$ &generic (information) relative entropy function (--)\\
$I_\nu$, $L_\nu$ &energy radiance (W m$^{-2}$ s sr$^{-1}$); entropy radiance (W K$^{-1}$ m$^{-2}$ s sr$^{-1}$) \\
${\vec j}_c$ &molar flux of chemical species $c$  (mol m$^{-2}$ s$^{-1}$)\\
${\vec j}_Q$, ${\vec j}_E$ &heat flux; energy flux (J m$^{-2}$ s$^{-1}$)\\
${\vec j}_S, {\vec J}_S$ &non-fluid entropy flux; total entropy flux (J K$^{-1}$ m$^{-2}$ s$^{-1}$) \\
%$k$ &Boltzmann constant(1.381 $\times$ 10$^{-23}$ J K$^{-1}$)\\
$k$, $k_{SB}$ & Boltzmann constant (J K$^{-1}$); Stefan-Boltzmann constant (W m$^{-2}$ K$^{-4}$)\\
$\K$ &steady-state flow constant (J K$^{-1}$ m$^{-3}$ s$^{-1}$) \\
%$L$ &Lagrangian function (--)\\ 
$m$ &fluid mass (kg)\\
$\vec{m}$, $\vec{n}$ &unit normal to area element; outward unit normal to control surface (-)\\
$M_c$ &molar mass of chemical species $c$ (kg mol$^{-1}$)\\
$n_c$ &molar density of chemical species $c$ (mol kg$^{-1}$)\\
$n_i$, $N$ &number of elements (balls) in partition $i$; total number of elements (--) \\
$p_i$, $q_i$ &inferred probability, prior probability (--) \\
$P$ &absolute pressure (Pa) \\
%$\mathbb{P}$ &governing probability distribution (--) \\
%$q_i$ &prior or source probability (--) \\
$R$ &number of constraints (--)\\
%$Re$ & Reynolds number (--)\\
%$s$  &specific thermodynamic entropy (J K$^{-1}$ kg$^{-1}$)\\
$S$, $\hat{S}$, $s$ &thermodynamic entropy (J K$^{-1}$); entropy per volume (J K$^{-1}$ m$^{-3}$); specific entropy (J K$^{-1}$ kg$^{-1}$)\\
$t$ &time (s)\\
$T$, $T_\nu$ &absolute temperature (K); radiative temperature (K)\\
$U$, $\hat{U}$, $u$ &internal energy (J); internal energy per volume (J m$^{-3}$); specific internal energy (J kg$^{-1}$) \\
${\vec v}$ &mass-average velocity vector (m s$^{-1}$)  \\
$V$ &volume (m$^{-3}$)\\
$\vec{x}$ &position vector (m) \\
$Z$ &partition function (--)\\ 
\hline\noalign{\smallskip}
\multicolumn{2}{l}{\textbf{Greek symbols} }\\
$\gamma$ &degeneracy of state (--)\\
%$\vec{\delta}$  &Kronecker delta tensor (--)\\
%$\epsilon$  &permutation tensor (--) or energy level (J) \\
$\epsilon$  & energy level (J) \\
$\lambda$, $\zeta$  &Lagrangian multiplier (various)\\
$\mu_c$ &molar chemical potential of species $c$ (J mol$^{-1}$)\\
$\nu$ &frequency of radiation (s$^{-1}$)\\
$\chi_{cd}$ &stoichiometric coefficient of species $c$ in the $d$th reaction (mol mol$^{-1}$)\\
$\hat{\dot{\xi}}_{c}$, $\hat{\dot{\xi}}_{d}$ &rate per volume of species $c$; of chemical reaction $d$ (mol m$^{-3}$ s$^{-1}$)\\
$\rho$ &fluid density (kg m$^{-3}$)\\
%$\rho_c$ &mass density of chemical species $c$ (kg kg$^{-1}$)\\
${\sigma}$ &amount of thermodynamic entropy produced (J K$^{-1}$)\\
$\dot{\sigma}$, $\hat{\dot{\sigma}}$, $\EParea$ &rate of thermodynamic entropy production (J K$^{-1}$ s$^{-1}$); rate per volume (J K$^{-1}$ m$^{-3}$ s$^{-1}$); rate per area (J K$^{-1}$ m$^{-2}$ s$^{-1}$)\\
%$\hat{\dot{\sigma}}$ &rate of thermodynamic entropy production per volume (J K$^{-1}$ m$^{-3}$ s$^{-1}$) \\
%$\EParea$ &rate of thermodynamic entropy production per area (J K$^{-1}$ m$^{-2}$ s$^{-1}$) \\
${\tens {\tau}}$ &viscous stress tensor (Pa) \\
$\Phi$, $\phi$ &potential (negative Massieu) function (--); Planck potential (J K$^{-1}$) \\ %$=-\lambda_0 - \ln Z$ (--)\\
%$\psi$ &overall mass-weighted body force potential (s$^{-2}$) \\
$\psi_c$ &mass-weighted body force potential on species $c$ (s$^{-2}$) \\
$\Omega$ &solid angle (sr) \\
$\halfEParea$ &rate of entropy production on one side of area (J K$^{-1}$ m$^{-2}$ s$^{-1}$) \\
\hline\noalign{\smallskip}
%\pagebreak  %FORCED PAGE BREAK; MAY NEED TO BE RESET AND \hline REINSERTED
\multicolumn{2}{l}{{\textbf{Superscripts, Subscripts and Indices}}}\\
$^*$  &stationary state \\
$^+, ^-$ &final, initial \\
%%
%\hline\noalign{\smallskip}
%\multicolumn{2}{l}{\textbf{Subscripts} }\\
$c, d$ &chemical species index, chemical reaction index \\
{$\path, \paths$} &{thermodynamic path index, set of allowable paths} \\
{$eq$, $st$} &{equilibrium system, steady-state system} \\
$f, nf, tot$ &fluid, non-fluid, total \\
$m, \nu$ &material, radiative \\
$i, j, k, \vec{i}$ &state indices \\
$in, out$ &in or out of control volume \\
$\ell, r$ &constraint indices \\
$\alpha, \beta$ &compartment indices\\
$\kappa$ &compartment boundary index \\
\hline\noalign{\smallskip}
\multicolumn{2}{l}{\textbf{Mathematical Symbols} }\\
$\overline{f}$, $\widetilde{f}$, $f'$ &time mean; ensemble mean; fluctuating component \\
$\dot{f}$, $\hat{f}$, $\breve{f}$, $\tilde{f}$ &per unit time; per unit volume; per unit area; per mole \\
$\langle f \rangle$ &expectation \\
$\lmean f \rmean$, $\lf f \rf$ &in-the-mean (product of means) form; mean fluctuating component \\ % of $f$ (based on products of expectations) \\
%$f'$ &fluctuating component of $f$ \\
%$\lf f \rf$ &mean fluctuating component of $f$
\end{longtable}
}

%\pagebreak

\end{document}